\documentclass[10pt,a4paper,twocolumn,floatfix,groupedaddress,superscriptaddress,prb,aps,showpacs]{revtex4-1}

\usepackage{graphicx}
\usepackage{amsmath}
\usepackage{amssymb}
\usepackage{amsfonts}
\usepackage{mathtools}
\usepackage{dcolumn}
\usepackage{bbm}
\usepackage{nicefrac}
\usepackage{pifont}

\newcommand{\RM}[1]{\MakeUppercase{\romannumeral #1{}}}

\usepackage{color}
\usepackage{bbding}

\begin{document}
\title{Minimal model for the frustrated spin ladder system BiCu$_2$PO$_6$}

\author{Leanna Splinter}
\email{leanna.splinter@tu-dortmund.de}
\affiliation{Lehrstuhl f\"{u}r Theoretische Physik I, 
Technische Universit\"{a}t Dortmund,
 Otto-Hahn Stra\ss{}e 4, 44221 Dortmund, Germany}

\author{Nils A. Drescher}

\author{Holger Krull}

\author{G\"otz S. Uhrig}
\email{goetz.uhrig@tu-dortmund.de}
\affiliation{Lehrstuhl f\"{u}r Theoretische Physik I, 
Technische Universit\"{a}t Dortmund,
 Otto-Hahn Stra\ss{}e 4, 44221 Dortmund, Germany}

\date{\textrm{September 5, 2016}}

\begin{abstract} 
To establish the microscopic model of the compound BiCu$_2$PO$_6$ is a challenging task. Inelastic neutron scattering experiments showed that the dispersion of 
this material is non-degenerate suggesting the existence of anisotropic interactions.
Here we present a quantitative description of the excitation spectrum for BiCu$_2$PO$_6$ 
on the one-particle level. The solution of the isotropic frustrated spin ladder by continuous
 unitary transformations is the starting point of our approach. 
Further couplings such as isotropic interladder couplings and anisotropic
interactions are included on the mean-field level.
Our aim is to establish a minimal model built on the symmetry allowed interactions 
and to find a set of parameters, which allow us to describe 
the low-energy part of the dispersion without assuming unrealistic couplings. 
\end{abstract}

\maketitle


\section{Introduction}
\label{chap_introduction}

In general, the interaction between two  spins in a quantum magnet is not completely isotropic 
due to  the fact that no crystallographic environment is entirely isotropic. As a consequence,
anisotropic interactions have to be considered in order to describe the properties of a compound
in an embracing quantitative way. 

Recently, Romh{\'a}nyi \textit{et al.} \cite{romha11,romha15} showed that small anisotropic 
interactions in SrCu$_2$(BO$_3$)$_2$, essentially a realization of the Shastry-Sutherland model
\cite{miyah99,knett00b,miyah03,knett04a}, 
give rise to non-trivial topological properties of the excitation spectrum and the phase diagram.
 In the compound (C$_7$H$_{10}$N$_2$)$_2$CuBr$_4$ (DIMPY) anisotropic interactions also exist and 
have the effect of lifting the triplet excitation degeneracy as well as broadening of the 
lines in electron spin resonance (ESR) \cite{glazk15}. These results  attracted great attention 
to the field of anisotropic interactions in low-dimensional spin systems in  experiment and 
in theory.

The anisotropic interaction, referred to as the Dzyaloshinskii-Moriya-interaction
(DM interaction) \cite{dzyal58,moriy60a,moriy60b} arises from the spin-orbit coupling (SOC)
which constitutes a relativistic correction to the non-relativistic description of atoms.
Thus it is particularly pronounced for elements with large atomic number implying 
a strong Coulomb potentials and high electronic velocities. The DM interaction 
between two localized spins $\mathbf{S}_{i}$ and $\mathbf{S}_{j}$
describes an antisymmetric interaction \cite{moriy60b} 
\begin{equation}
\mathcal{H}_{\mathbf{D}}=\mathbf{D}_{ij}\left(\mathbf{S}_{i}\times\mathbf{S}_{j}\right), 
\end{equation}
which arises already in linear order in the SOC. Additionally a symmetric anisotropic exchange 
\begin{equation}
\mathcal{H}_{\Gamma}=\sum_{\alpha,\beta} 
\Gamma_{ij}^{\alpha\beta}S_{i}^{\alpha}S_{j}^{\beta} 
\end{equation}
occurs from the SOC, which is of quadratic order in the SOC; $\alpha$ and $\beta$ label the spin components.
In spite of being quadratic in the SOC, the symmetric terms are not negligible 
\cite{shekh92} compared to the antisymmetric ones.

Another candidate for important DM interaction is the compound BiCu$_2$PO$_6$ (BCPO) 
which received  much attention in the last decade \cite{koham12,wang10c,casol10,kotes10,sugim13,
koham14,casol13,alexa10,bobro09,plumb13,mentr09,sugim14,sugim15,lavar11,kotes07,mentr06,nagas14}.
It is difficult to estimate the relevance of the SOC. Although
bismuth has a large atomic number (Z=83) it does not host the localized 
spin which resides at the copper ions. Thus the DM interactions depend on the details of
the super exchange paths and to what extent the bismuth ions are involved or not.

BCPO is a realization of a spin ladder in the intermediate energy range \cite{wang10c} 
($J\sim$10\,meV) what makes it an interesting material to analyze on the theoretical and on the 
experimental side. Its crystallographic structure contains tube-like, frustrated spin-$1/2$ Heisenberg ladders. These spin ladders are coupled among one another in one spatial direction, 
which makes BCPO  a two-dimensional material \cite{wang10c,mentr06}. 
The actual ladder structure of BCPO is still controversial and has been a point of 
argument in the past \cite{kotes07,mentr09}.

Several properties of BCPO have been measured in the last years, such as field-induced phase 
transitions \cite{koham12}, the thermal conductivity \cite{nagas14}, the magnetic susceptibility 
\cite{kotes07}, the heat capacity \cite{kotes07} and the spin excitation spectrum 
\cite{plumb14,plumb15}. Even the effects of doping BCPO with Zn or Ni on the Cu site \cite{kotes10} 
and V on the P site \cite{mentr06}  have been analyzed.

On the theoretical side various methods, such as the density matrix renormalization group (DMRG) 
\cite{sugim15,lavar11,sugim13,tsirl10}, quantum Monte Carlo  simulations (QMC) 
\cite{alexa10,casol10}, exact diagonalization (ED) \cite{lavar11,tsirl10}, 
density-functional calculations of the band structure \cite{kotes07,mentr09,tsirl10} and 
quadratic bond operator theory \cite{plumb14,plumb15} have been used to describe the magnetic properties.

A recent theoretical analysis argued that the DM interactions in BCPO are as large as 
$D\approx 0.6J$ where $J$ is the isotropic Heisenberg exchange of the corresponding bond \cite{plumb14,hwang16a}. 
Lately these values were revised \cite{plumb15,hwang16a} to $D\approx 0.3J$. 
The analysis suggesting the lower relative values includes the effects of the interaction of the elementary 
excitations, i.e., triplons.

Keeping in mind that DM interactions arise from the SOC we classify these 
values as extremely large. A standard estimate for the relative strength
of $D/J$ is $|\Delta g|/g$ where $g$ is the gyromagnetic ratio $g\approx2$ and
$\Delta g= g-2$. For spins in copper ions $\Delta g$ varies from zero to $0.4$ so 
that any value of $D/J$ beyond $0.2$ must be considered remarkable.
Thus it is our motivation to derive a quantitative one-particle description for the low-lying magnetic excitation modes of BCPO within a minimal spin model.
In particular, we want to investigate which values of the DM interactions
are required to describe the magnetism in BCPO.

This article is set up as follows. First, we present the structure of BCPO and discuss the controversial point concerning the ladder structure briefly. In the next section, we start with a brief overview of the method of continuous unitary transformations
which constitutes the basis for our calculations. 
After that we present the starting point for our calculations 
and the choice of parameters for the isotropic model. 
In Sect.\ \ref{chap_Dcomponents} the directions of the $\mathbf{D}$-vectors of the 
DM interactions are determined examplarily. The relation between the $\mathbf{D}$-components 
and the matrix elements of the symmetric tensor $\Gamma$ are derived by mapping the anisotropic interactions between two spins onto a pure isotropic interaction in a rotated basis. 
The perturbative method used to compute the influence of the anisotropic interactions on the
dispersion of BCPO is illustrated in Sect. \ref{method}. In the following section, the results 
are discussed. As a consequence of these results, we propose an modification of the next-nearest neighbor interaction $J_{2}$ to improve agreement between experiment and theory in Sect.\ 
\ref{sect_alternatingJ2} and compare its results to the previous ones. 
We obtain a considerably improved set of parameters.
Finally, we conclude our study  in Sect. \ref{summary} including an outlook. 


\section{Structure of BCPO}
\label{chap_structure}

We focus only on the spin model of BCPO and refer to Tsirlin \textit{et al.} 
\cite{tsirl10} for a detailed 
description of the crystal structure including the spatial arrangement of the
relevant ions. 
The magnetic structure of BCPO is dominated by tube-like arranged spin ladders coupled among
themselves leading to a two-dimensional lattice \cite{wang10c,mentr06}. 
The tubes in  BCPO constitute frustrated spin ladders which are formed by two 
crystallographically different types of copper ions. 

\begin{figure}[htb]
\centering
\includegraphics[scale=0.16]{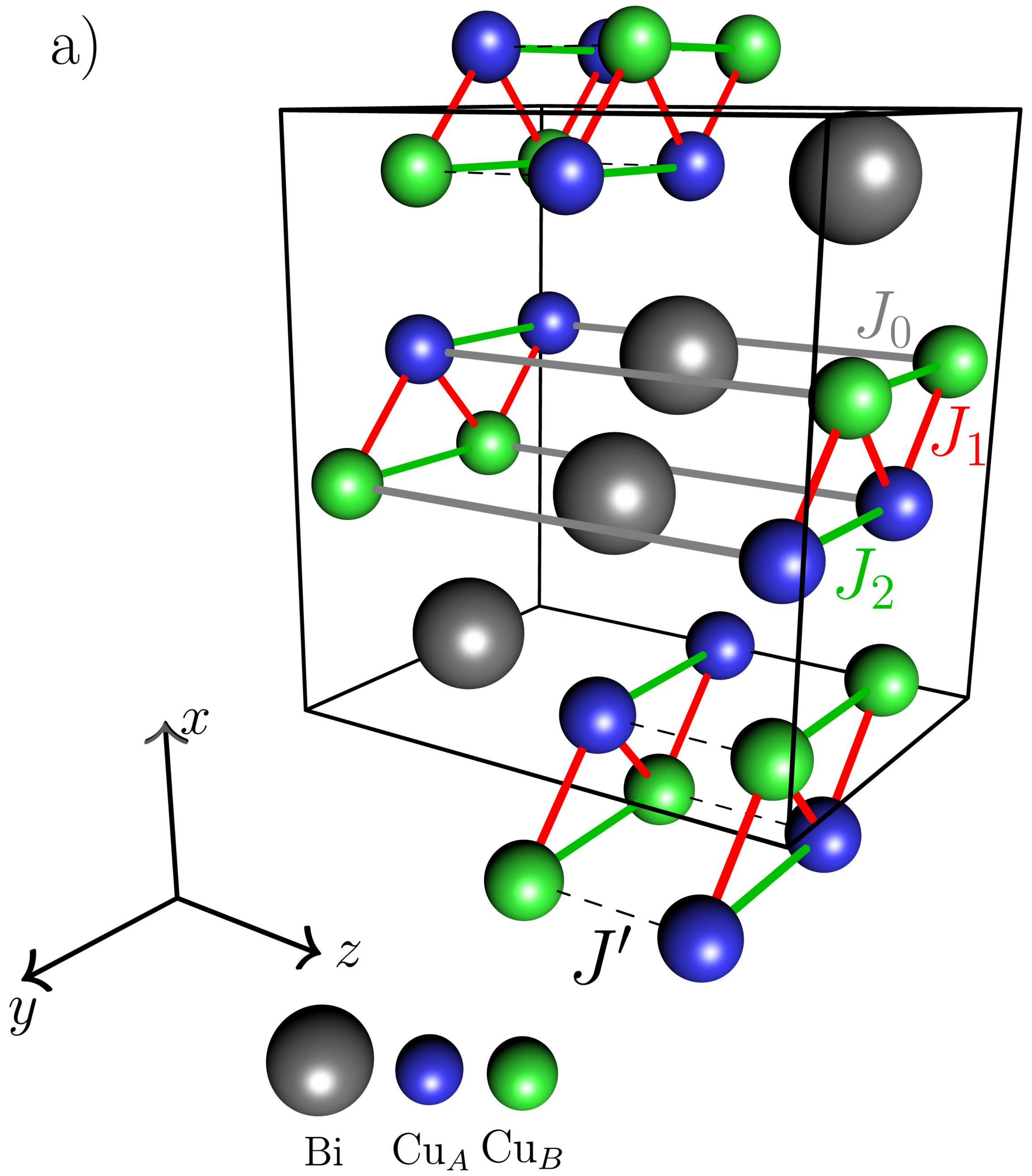}
\includegraphics[scale=0.16]{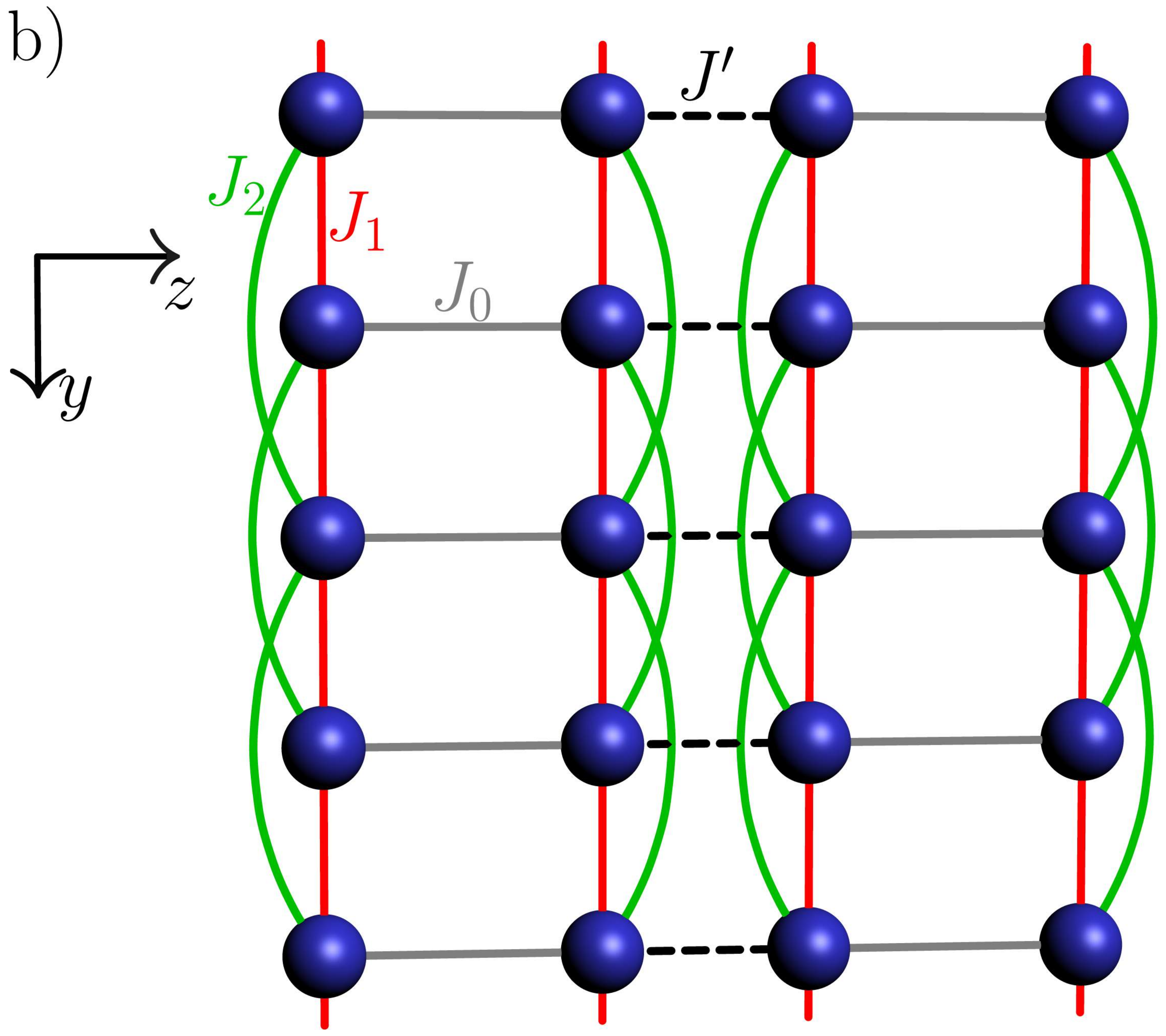}
\caption{a) Crystal structure of BCPO. The unit cell is orthorhombic and contains coupled frustrated 
spin ladders formed by the two inequivalent copper ions Cu$_{A}$ and Cu$_{B}$. 
We omitted the phosphorus and oxygen ions for a better overview.
b) Effective spin model. The analyzed model is made of frustrated spin ladders, which are coupled by an interladder 
coupling $J^{\prime}$. The inequivalence of the copper ions is neglected.}
\label{pic_structure_bcpo}
\end{figure}

The two types of copper ions Cu$_{A}$ and Cu$_{B}$ alternate along the ladder in $y$-direction as
 shown in Fig.\ \ref{pic_structure_bcpo}. 
The coupling in the $xy$-plane between the spins belonging to different types of copper ions,  
constitutes the nearest neighbor (NN) interaction $J_{1}$ and forms a 
zigzag pattern. The couplings in $z$-direction are labelled $J_{0}$ and $J^{\prime}$ and 
act also between copper ions of different types. 
It is reasonable to assume a difference between the next-nearest neighbor (NNN) couplings 
$J_{2}$ and $J_{2}^{\prime}$ which couples the copper ions of the same 
type (Cu$_{A}$-Cu$_{A}$ and Cu$_{B}$-Cu$_{B}$) in $y$-direction \cite{tsirl10}. 
First, we neglect the difference between $J_{2}$ and $J_{2}^{\prime}$ and denote the 
NNN interaction by $J_{2}$. In Sect.\ \ref{sect_alternatingJ2} we come back to this point 
discussing various extensions.

Considering the couplings $J_{0}$ and $J^{\prime}$ in $z$-direction it is not clear which of them
describes the rung coupling of the spin ladder and which the interladder coupling. 
The crystal structure is consistent with both options. 
Koteswararao \textit{et al.} \cite{kotes07} proposed $J^{\prime}$ to be the rung coupling 
due to the shorter distance between the concerned copper ions making stronger super
exchange possible. Then $J_{0}$ was identified as the interladder coupling. 
The basis for this assignment were band structure calculations and measured susceptibility data. 

In return, Mentr\'e \textit{et al.} \cite{mentr09} 
suggested $J^{\prime}$ to be the  interladder coupling and $J_{0}$ to be the rung coupling of the ladders. Their arguments for this assignment were based on the angles of 
the associated bonds, band structure calculations and inelastic neutron scattering (INS) measurements. Plumb \textit{et al.} \cite{plumb13} 
verified Mentr\'e's proposal by analyzing the intensity modulation along the $x$- and 
$z$- direction. For this reason, we use the assignment suggested by Mentr\'e. But we stress that our results for the dispersion do not depend on the assignment 
between $J^{\prime}$ and $J_{0}$. This is the case because we do not address spectral weights in the present article.

Another issue was the question whether BCPO has to be described by a one-dimensional or 
by a two-dimensional model. For the answer one has to compare the 
value of the interladder coupling $J^{\prime}$ 
with the intraladder couplings $J_{0}$, $J_{1}$ and $J_{2}$. 
In the $yz$-plane, noticeable dispersions exist and 
therefore an interladder coupling has to be taken into 
account in order to capture the essential features of BCPO \cite{tsirl10,plumb13}. 
The dispersion along the $x$-direction is hardly detectable and can be neglected 
\cite{tsirl10,plumb13}. As a result, BCPO can be described as a 
two-dimensional frustrated spin ladder system  with an interladder coupling 
$J^{\prime}$ in $z$-direction. Due to the absence of inversion symmetry about the center of the Cu-Cu bonds, see 
Fig.\ \ref{pic_structure_bcpo}, anisotropic interactions may occur in BCPO \cite{moriy60b}. 
Other observations, such as discrepancies between 
measured gap values and the calculated ones excluding anisotropic interactions \cite{tsirl10}, 
indicate that anisotropic interactions must be present. In addition, the difference between the
gap value from neutron-scattering 
and from thermodynamic measurements underlines that strong anisotropic interactions are
required to receive a comprehensive understanding of BCPO \cite{plumb13}.


\section{The isotropic spin ladder}
\label{chap_isotropic}

Here, we present the results for the isotropic ladder. They constitute the starting point
of our study because the anisotropic couplings are expected to be small relative to the
isotropic ones. In order to provide a self-contained study we give a brief overview about 
the method employed, i.e., continuous unitary transformations.

\subsection{Continuous unitary transformations}
\label{chap_CUT}

With the help of continuous unitary transformations (CUTs) it is possible to 
derive effective models $\mathcal{H}^{\text{eff}}$ from complex initial systems $\mathcal{H}$
in a systematic and controlled way. 
The main idea of CUTs is to simplify $\mathcal{H}$ step by step 
by applying unitary transformations. Its basic concept has been introduced by Wegner 
\cite{wegne94} and  by Glazek and Wilson \cite{glaze93,glaze94}, for a review see
Ref.\ \onlinecite{kehre06}.

Instead of a discrete unitary transformation the CUT approach
 uses continuous unitary transformations $U\left(l\right)$, which depend on 
the so-called flow parameter $l$. Therefore the relation 
\begin{equation}
\mathcal{H}\left(l\right)=U\left(l\right)\mathcal{H}U^{\dagger}\left(l\right) 
\end{equation}
holds with the starting condition $U\left(0\right)=\mathbbm{1}$.
The flow equation of the Hamiltonian is defined by the differential equation
\begin{equation}
\label{flow_hamiltonian}
\partial_{l}\mathcal{H}\left(l\right)=\left[\eta\left(l\right),\mathcal{H}\left(l\right)\right].
\end{equation}
Here the anti-hermitian generator 
$\eta\left(l\right)=\left(\partial_{l}U\left(l\right)\right)U^{\dagger}\left(l\right)$ of the CUT is introduced. Equation \eqref{flow_hamiltonian} can be interpreted 
as a system of coupled differential equations 
for the prefactors of the operators, which occur in the 
Hamiltonian $\mathcal{H}\left(l\right)$.

In general, an infinite number of differential equations ensues which need to be solved.
Thus one has to define an appropriate truncation scheme. A truncation scheme limits 
the terms in $H\left(l\right)$ to ensure a sufficient good description of $H\left(l\right)$.
The various types of CUTs differ in the employed truncation scheme. 
We used the directly evaluated enhanced perturbative CUT (deepCUT) introduced
four years ago \cite{krull12}. In this scheme, operators and terms in the
differential equations are kept or omitted according to 
their effect in powers of the expansion parameter $x$ on certain
target quantities. In the present study, the target quantity is the
dispersion of the triplons. 
If $n$ denotes the order up to which the target quantitiy should be computed, all operators 
and terms are kept which affect the target
quantity  in the order $m\leq n$ in $x$.

In the limit $l \to \infty$ the effective Hamiltonian 
\begin{equation}
\mathcal{H}^{\text{eff}}=U\left(\infty\right)HU^{\dagger}\left(\infty\right)
\end{equation}
is obtained and can be analyzed.

In essence, a CUT is a change of basis. This means that observables $O$ are also mapped onto effective observables $O^{\text{eff}}$ using the same unitary transformations. For their transform one obtains an analogous set of coupled differential 
equations from
\begin{equation}
\partial_{l}O\left(l\right)=\left[\eta\left(l\right),O\left(l\right)\right].
\end{equation}
In the limit $l \to \infty$, we obtain the effective observable $O^{\text{eff}}$.

The generator $\eta\left(l\right)$ determines the flow of the Hamiltonian, see Eq.\ \ref{flow_hamiltonian}. There is a variety of generators which have slightly different 
properties. For our problem we used the 1n-generator \cite{fisch10} which reads
\begin{equation}
\eta_{1\text{n}}\left(l\right)=\mathcal{H}_{0}^{+}\left(l\right)+\mathcal{H}_{1}^{+}\left(l\right)
-\mathcal{H}_{0}^{-}\left(l\right)-\mathcal{H}_{1}^{-}\left(l\right).
\end{equation}
The operators $\mathcal{H}_{0}^{+}\left(l\right)$ and $\mathcal{H}_{1}^{+}\left(l\right)$ contain all terms of $\mathcal{H}\left(l\right)$ 
which create more quasiparticle than they annihilate out of states 
with 0 and 1 quasiparticle at least. In return, the operators 
$\mathcal{H}_{0}^{-}\left(l\right)$ and $\mathcal{H}_{1}^{-}\left(l\right)$ refer to all terms of 
$\mathcal{H}\left(l\right)$ annihilating more 
quasiparticles than creating. Clearly, $\mathcal{H}_{m}^{-}\left(l\right)$
is the hermitian conjugate of $\mathcal{H}_{m}^{+}\left(l\right)$.
The 1n-generator decouples the subspaces containing 
zero and one quasi-particle from all other subspaces. 
Thus this generator is particularly suited to compute
 the ground-state energy and the dispersion \cite{fisch10}.

\subsection{Results for the isotropic spin ladder}
\label{isotropic}

The first step to describe the measured dispersion of BCPO is to analyze 
the spectrum of a single frustrated isotropic spin ladder with the Hamiltonian
\begin{subequations}
\label{hamiltonian_isotropic}
\begin{align}
\mathcal{H}_{\text{ladder}}&=J_{0}\mathcal{H}_{0}+J_{1}\mathcal{H}_{1}+J_{2}\mathcal{H}_{2}\\
\mathcal{H}_{0}&=\sum_{i}\mathbf{S}_{i}^{\mathrm{L}}\mathbf{S}_{i}^{\mathrm{R}}\\
\mathcal{H}_{1}&=\sum_{i,\tau}\mathbf{S}_{i}^{\tau}\mathbf{S}_{i+1}^{\tau}\\
\mathcal{H}_{2}&=\sum_{i,\tau}\mathbf{S}_{i}^{\tau}\mathbf{S}_{i+2}^{\tau},
\end{align}
\end{subequations}
where $i$ is the rung index. The variable $\tau$ assumes the values $\mathrm{L}$ for the left leg
 of the spin ladder and $\mathrm{R}$ for the 
right leg. We define the ratios $x=\nicefrac{J_{1}}{J_{0}}$ and $y=\nicefrac{J_{2}}{J_{1}}$.
The parameter $x$ is the expansion parameter around the limit of decoupled rungs, i.e., in the
limit $x\to 0$ at constant  $y$ no interdimer coupling is left. So $x$ is used in 
the deepCUT approach as th parameter defining the truncation scheme.
The paramter  $y$ controls the relative strength of the NN and NNN coupling along
the legs of the ladder. 

Because the structure of BCPO consists of frustrated spin ladders coupled by an interladder coupling $J^{\prime}$, it is necessary to take the effect of $J^{\prime}$ into account 
as well. To this end, we start from  the effective model of a single frustrated 
spin ladder obtained by  deepCUT as sketched above. That means we consider the 
following Hamiltonian of dispersive triplons
\begin{equation}
\label{hamiltonian_isotropic_solved}
\mathcal{H}_{\text{ladder}}^{\text{eff}}=\sum_{k,\alpha}\omega_{0}
\left(k\right)t_{k}^{\alpha,\dagger}t_{k}^{\alpha}.
\end{equation}
The operator $t_{k}^{\alpha,\dagger}$ $\left(t_{k}^{\alpha}\right)$ 
creates (annihilates) a triplon \cite{sachd90,schmi03c} with momentum 
$k$ and flavor $\alpha\in\{x,y,z\}$. The dispersion of a single frustrated spin ladder is denoted with $\omega_{0}\left(k\right)$. Possible interactions between two or even more triplons
are left out at this stage because we do not have experimental indications for their
relevance.

Next, we also transform other operators to their effective
counter parts by the same CUT. In particular, we need 
the spin operator $S_{i}^{\alpha,\mathrm{R}}$ expressed in triplon 
operators
\begin{equation}
\label{effective_spinoperator}
S_{i,\text{eff}}^{\alpha,\mathrm{R}}=\sum\limits_{\delta=-n}^{n}a_{\delta}
\left(t_{i+\delta}^{\alpha,\dagger}+t_{i+\delta}^{\alpha}\right)+ \ldots .
\end{equation}
The  dots refer to omitted terms of  normal-ordered bilinear terms and terms of even higher
number of triplon operators which we neglect for our calculations similar to previous
applications \cite{uhrig04}. The index $\delta$ runs from $-n$ to $n$ in integer steps
while $n$ denotes the order up to which the spin ladder was solved by the CUT. The effective spin operator \eqref{effective_spinoperator} is not local any more, but a superposition of triplon operators from rung $i-n$ to rung $i+n$. The coefficients $a_{\delta}$ indicate the 
probability amplitude of the triplon operator on rung $i+\delta$.
Physically, this expresses the fact that the initial triplon which is completely
local becomes smeared out when the effect of the interrung couplings $J_1$ and $J_2$
are considered.

We only focus on the linear terms in the effective spin operators. On this level
of description, the relation
\begin{equation}
\label{eq:leftright}
S_{i,\text{eff}}^{\alpha,\mathrm{R}}=-S_{i,\text{eff}}^{\alpha,\mathrm{L}}
\end{equation}
is valid. It is based on the fact that triplon excitations have odd parity relative to the ground
state with respect to reflection on the center line of the spin ladder \cite{sachd90,schmi05b}, see also symmetry S$_{xy}$ in section \ref{symmetriesDcomponents}.

From now on, we treat the triplons as free bosons in a mean-field approach. This
approach constitutes an approximation, but it is justified by the relative smallness of the interladder coupling $|J'/J_0|\ll 1$. 
The Fourier transformation of \eqref{effective_spinoperator} yields 
\begin{equation}
\label{eq:effspin}
S_{\text{eff}}\left(k\right)^{\alpha,\mathrm{R}}=a\left(k\right)\left(t_{k}^{\alpha,\dagger}+
t_{-k}^{\alpha}\right)
\end{equation}
using the quantity
\begin{subequations}
\label{a(k)}
\begin{align}
a\left(k\right) &=\sum_{\delta}a_{\delta}\mathrm{e}^{\mathrm{i}k\delta}
\\
& = \sum_{\delta}a_{\delta}\cos\left(k\delta\right).
\label{a(k)2}
\end{align}
\end{subequations}
The  absolute value squared of $a\left(k\right)$ corresponds to the 
weight of the dominant single-particle mode in 
the dynamic  structure factor at zero temperature under the made assumptions.
In the single mode approximation this weight equals the momentum 
resolved static structure factor.
The Eq.\ \eqref{a(k)2} is valid because the spin ladder 
fulfills the relation $a_{\delta}=a_{-\delta}$ due to the mirror symmetry about a rung, see symmetry S$_{xz}$ in Sect.\ \ref{symmetriesDcomponents}.

The Hamiltonian 
\begin{equation}
 \mathcal{H}^{\prime}=J^{\prime}\sum_{i,j}\mathbf{S}_{i,j}^{\mathrm{R}}
\mathbf{S}_{i,j+1}^{\mathrm{L}}
\end{equation}
describes the coupling between two adjacent spin ladders with the coupling strength
 $J^{\prime}$. The index $i$ denotes the rung again and the index $j$ 
labels the ladder. Using the effective operators from Eq.\ \eqref{effective_spinoperator} in 
Fourier transformed form leads to the 
effective Hamiltonian of the interladder coupling 
\begin{equation}
\label{hamiltonian_2D}
\mathcal{H}^{\prime,\text{eff}}=
-J^{\prime}\sum_{k,l,\alpha}d_{k,l}\left(t_{k,l}^{\alpha,\dagger}+t_{-k,-l}^{\alpha}\right)
\left(t_{k,l}^{\alpha}+t_{-k,-l}^{\alpha,\dagger}\right)
\end{equation}
with the abbreviation
\begin{equation}
 d_{k,l}=\cos\left(2\pi l\right)a^2\left(k\right).
\end{equation}
Here the variable $l$ indicates the wave vector perpendicular to the spin ladder (in $z$-direction, see Fig. \ref{pic_structure_bcpo}) measured 
in reciprocal lattice units (r.l.u).

The complete Hamiltonian is the sum of $\mathcal{H}_{\text{ladder}}$ in 
Eq.\ \eqref{hamiltonian_isotropic_solved} 
for all ladders and of $\mathcal{H}^{\prime}$ in Eq.\ \eqref{hamiltonian_2D}. 
Since the interladder coupling is  weak 
compared to the ladder couplings $J_{0}$, $J_{1}$ and $J_{2}$ we use a standard Bogoliubov 
transformation to obtain the complete two-dimensional dispersion 
\begin{equation}
\omega\left(k\right)=\sqrt{\left(\omega_{0}\left(k\right)\right)^{2}-4J^{\prime}d_{k,l}\omega_{0}\left(k\right)}
\end{equation}
of the complete isotropic system.

\begin{figure}[htb]
\includegraphics[width=\columnwidth]{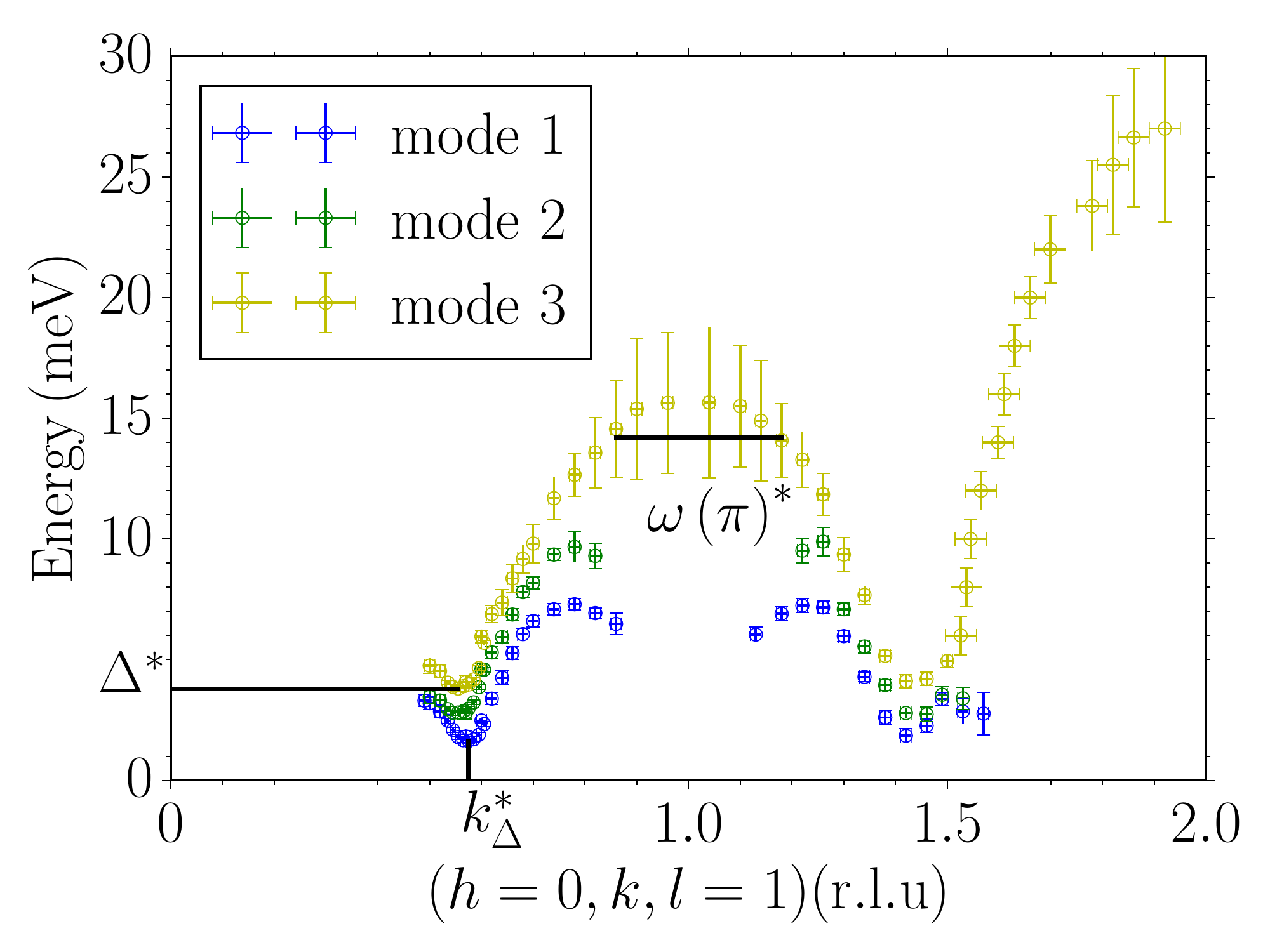}
\caption{Measured dispersion from Ref.\ \onlinecite{plumb14} and \onlinecite{plumb15}. 
The defined values for the selection criteria are marked as follows:
$k_{\Delta}^{*}=0.575\,\mathrm{\left(r.l.u\right)}$ describes the position of the gap of mode 1, 
$\omega\left(\pi\right)^{*}=14\,\text{meV}$ is the
average value of mode 3 at $k=\pi$ and $\Delta^{*}=3.8\,\text{meV}$ is the corresponding gap value.}
\label{measured_data}
\end{figure}

\begin{figure}[htb]
\includegraphics[width=\columnwidth]{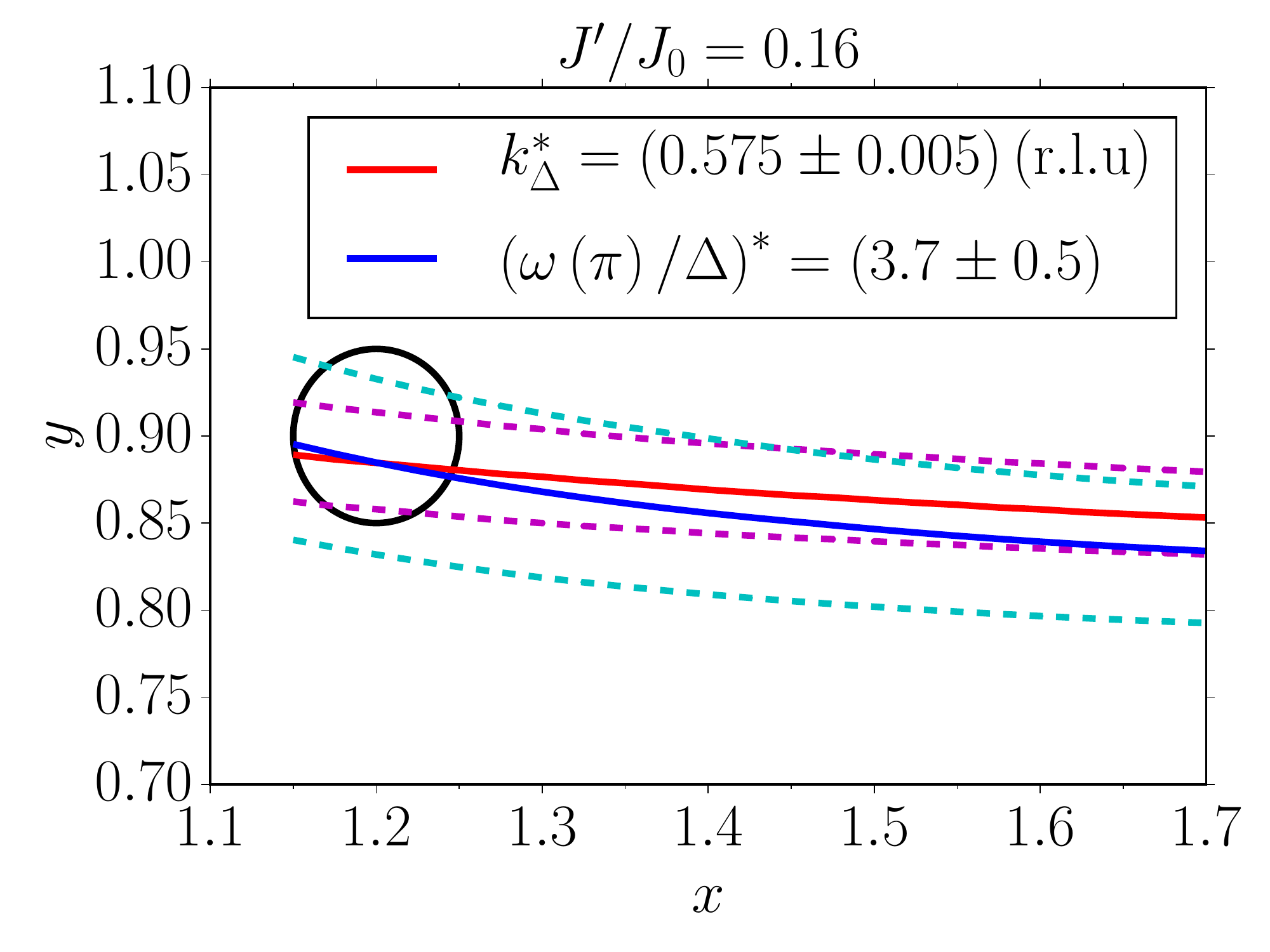}
\caption{Compliance with the selected criteria. 
The blue solid line is defined by  
$\omega\left(k\right)/\Delta=\left(\omega\left(\pi\right)/\Delta\right)^{*} = 3.7$;
the corresponding dashed lines indicate deviations by $\pm 0.5$, i.e., 
$\omega\left(k\right)/\Delta= 3.7\pm 0.5$.
The red line shows $k_{\Delta}=k_{\Delta}^{*}=0.575\,\mathrm{r.l.u.)}$; 
the corresponding dashed lines indicate deviations by $\pm 0.005$, i.e.,
$k_{\Delta}=\left(0.575\pm0.005\right)\,\mathrm{r.l.u.}$.
The circle marks the area where both criteria are fulfilled at about 
$x\approx 1.2$ and $y\approx 0.9$. 
The interladder coupling $J^{\prime}$ is set here to $J^{\prime}/J_{0}=0.16$.}
\label{isotropic_2D}
\end{figure}

It is not possible to describe the measured dispersion data of BCPO with an isotropic model completely because its dispersion is threefold degenerate unlike in experiment. 
Our aim in the analysis with the isotropic model is to find the best matching 
values of the parameters $x$ and $y$. 
To this end, we choose two features of the dispersion which are essential and
which should be described in the isotropic model. 
The first criterion is the $k$-value where  the gap $\Delta$ occurs.
We denote this value by $k_{\Delta}^{*}$. The second criterion is the ratio between 
the lower maximum $\omega\left(\pi\right)$ and the gap $\Delta$. 
Therefore, we analyze the first published results of the dispersion \cite{plumb14,plumb15}, see Fig.\ 
\ref{measured_data}.

Concerning  $k_{\Delta}^{*}$ we have to focus on one of the three measured modes and 
take its gap position as the desired value. 
We choose mode 1 for this criterion because it is the lowest lying mode. Its position is 
read off to be $k_{\Delta}^{*}=\left(0.575\pm0.005\right)\mathrm{(r.l.u)}$.

To identify a suitable value for the ratio $\left(\omega\left(\pi\right)/\Delta\right)^{*}$
is difficult because one has to guess which gap value the system would have
if the anisotropic couplings were not present. We decided to use mode 3 because
it appears to be the mode which can be followed through the whole Brillouin zone.
A posteriori, we will verify that this assignment makes sense because the additional
anisotropic couplings tend to reduce the dispersion in energy.
Because the values measured around $k=\pi$ have large error bars
we take the average of the values between
$k_{\text{start}}=0.8\,\mathrm{\left(r.l.u\right)}$ to 
$k_{\text{end}}=1.2\,\mathrm{\left(r.l.u\right)}$.
The rounded value finally used is  $\omega\left(\pi\right)^{*}=14\,\text{meV}$. 

Since we use mode 3 to read off a value for $\omega\left(\pi\right)^{*}$ 
we consistenly take the gap value of mode 3 to obtain the desired ratio
$\left(\omega\left(\pi\right)/\Delta\right)^{*}$. 
The gap value of mode 3 is $\Delta^{*}=3.8\,\text{meV}$ and thus we reach the ratio 
$\left(\omega\left(\pi\right)/\Delta\right)^{*}=3.7$ as reference. 
Due to the large error bars, we estimate that a deviation from this value of up to  0.5 
is still acceptable.

To find the best matching values of $x$ and $y$ we present 
the curves defined by  $k_{\Delta}=k_{\Delta}^{*}$ and 
$\omega\left(k\right)/\Delta=\left(\omega\left(\pi\right)/\Delta\right)^{*}$ 
in Fig.\ \ref{isotropic_2D} including the respective regions of acceptable deviations.
As one sees both criteria are fulfilled well
for $x\approx1.2$ and $y\approx0.9$. In this analysis, we used a relative interladder coupling 
of $J^{\prime}/J_{0}=0.16$ as done previously \cite{plumb14,plumb15}. As Fig.\ \ref{isotropic_2D} shows 
an overlap of both selection criteria in the tolerated error range for larger $x$ than 1.2, we compared the 
isotropic dispersion of larger $x$ with the measured dispersions. But analyzing the dispersion with values of $x=1.3$ to $x=1.7$ and $y=0.9$ 
does not show any improvement. Similarly, a variation of $y$ does not improve the results. 
Thus, the parameters $x\approx1.2$, $y\approx0.9$, and  $J^{\prime}/J_{0}=0.16$
define our starting point for the minimal isotropic model for BCPO. This will be refined
in the sequel.


\section{Analysis of the antisymmetric and symmetric anisotropic couplings}
\label{chap_Dcomponents}

Starting from the minimal isotropic model determined in the previous section,
we consider here anisotropic couplings, i.e., the Hamiltonian
\begin{equation}
 \mathcal{H}=\mathcal{H}_{\text{ladder}}+\sum_{i,j}\mathbf{D}_{ij}\left(\mathbf{S}_{i}\times\mathbf{S}_{j}\right)+
\sum_{i,j}\sum_{\alpha,\beta}\Gamma_{ij}^{\alpha\beta}S_{i}^{\alpha}S_{j}^{\beta}.
\end{equation}
It consists of  the isotropic spin ladder  
$\mathcal{H}_{\text{ladder}}$ \eqref{hamiltonian_isotropic} and 
the DM interactions 
$\mathbf{D}_{ij}\left(\mathbf{S}_{i}\times\mathbf{S}_{j}\right)$ 
and the symmetric anisotropic exchanges 
$\Gamma_{ij}^{\alpha\beta}S_{i}^{\alpha}S_{j}^{\beta}$. 
We want to stress that the sums with the indices $i$ and $j$ 
count each pair of spins only once.

\begin{figure}[htb]
\centering
\includegraphics[scale=0.2]{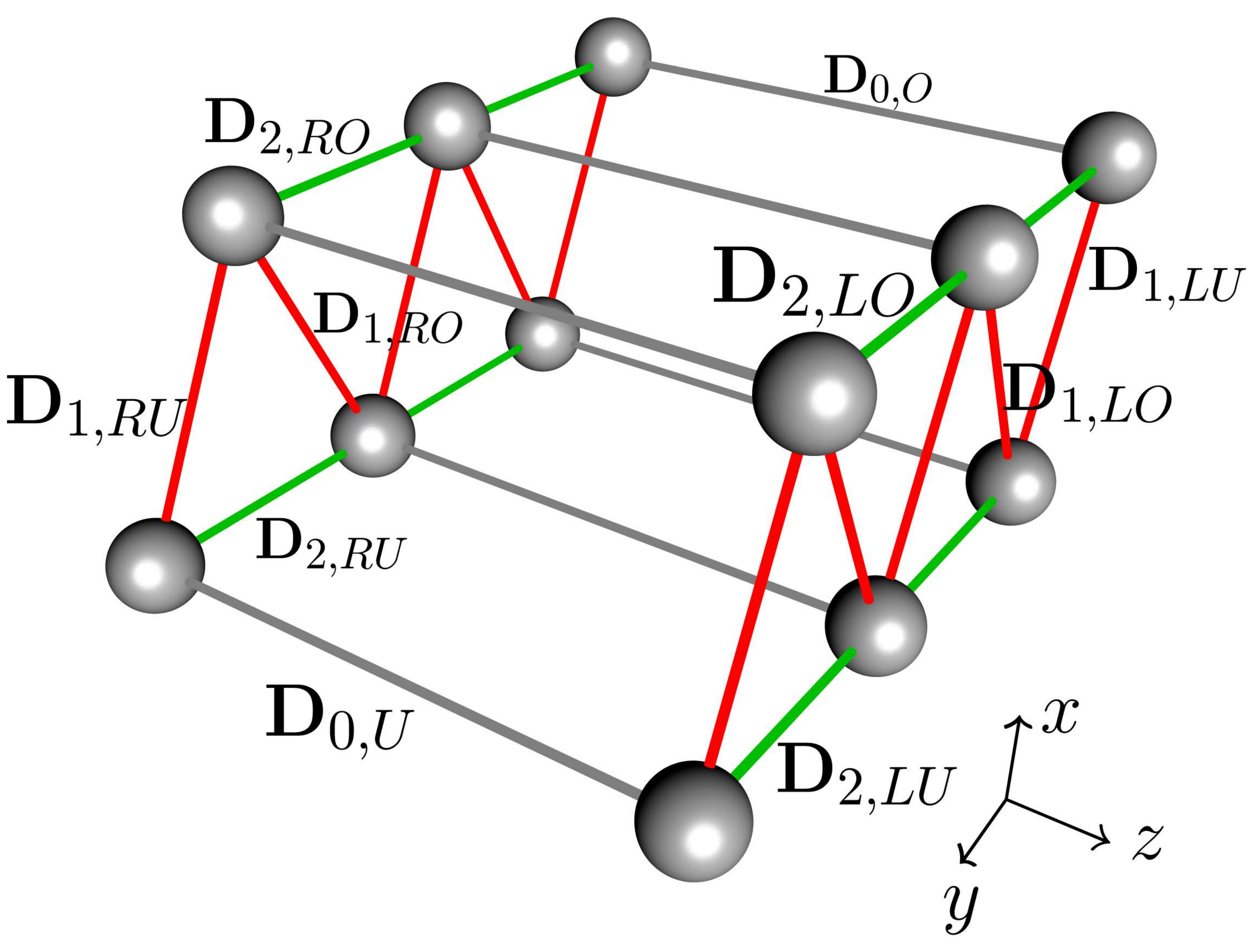}
\caption{Scheme of the spin ladder structure of BCPO. 
The gray spheres represent the copper ions Cu$_{A}$ and Cu$_{B}$, 
see Fig.\ \ref{pic_structure_bcpo}. The different bonds are labelled by
 the corresponding $\mathbf{D}$-vectors. The unit cell of the spin ladder 
contains an upper and a lower rung.}
\label{pic_spinladder_dm}
\end{figure}

We denote the couplings concerning the rungs of the ladder with the index 0, thus $J_{0}$, 
$\mathbf{D}_{0}$, and $\Gamma_{0}^{\alpha\beta}$. 
The couplings concerning the NN interactions are marked with the index 1, thus 
$J_{1}$, $\mathbf{D}_{1}$, and $\Gamma_{1}^{\alpha\beta}$. 
Finally, the components considering the NNN bonds carry the index 2, thus 
$J_{2}$, $\mathbf{D}_{2}$, and $\Gamma_{2}^{\alpha\beta}$,
see also Fig.\ \ref{pic_spinladder_dm}.

\subsection{Symmetries of the $\mathbf{D}$-components}
\label{symmetriesDcomponents}

First, we have to specify the direction of each DM-vector 
$\mathbf{D}_{ij}$. The components of $\mathbf{D}$ are real and constant coefficients. Each bond
on the ladder holds one $\mathbf{D}$ vector, see Fig.\ \ref{pic_spinladder_dm}.
The direction of the $\mathbf{D}$ vectors is restricted by the selection rules formulated by Moriya \cite{moriy60b}. To decide whether a 
component $D_{ij}^{\alpha}$ has to vanish or not, it is necessary to apply these rules to the crystal structure of BCPO, see Sect.\ \ref{chap_structure}. 
Our convention in the notation of the DM-vectors is the following.
For the NN and NNN bonds the spin operators in the outer product 
$\mathbf{D}_{ij}\left(\mathbf{S}_{i}\times\mathbf{S}_{j}\right)$ 
are ordered according to ascending $y$-coordinate. 
The convention for the rung couplings is to order the spin operators according to 
ascending $z$-coordinate.

We stress that in this symmetry analysis we do not distinguish between the two 
inequivalent copper sites Cu$_{A}$ and Cu$_{B}$, see Figs.\ 
\ref{pic_structure_bcpo} b) and \ref{pic_spinladder_dm}, but treat all sites as equal. 
The prevailing symmetries of the crystal structure are the following
\begin{itemize}
\item[1.]{RS$_{y}$: Rotation by $\pi$ about $\vec{y}$ located in the middle of the 
ladder tube and a shift by half a unit cell.}
\item[2.]{R$_{x}$: Rotation by $\pi$ about $\vec{x}$ located in the middle of a rung.}
\item[3.]{S$_{xy}$: Reflection at the $xy$-plane located in the middle of the ladder.}
\item[4.]{S$_{xz}$: Reflection at the $xz$-plane perpendicular through a rung.}
\item[5.]{SS$_{yz}$: Reflection at the $yz$-plane located in the middle of the ladder and a shift by half a unit cell.}
\end{itemize}
 
Next, we apply the above five symmetries to each bond, see Figs.\ 
\ref{pic_structure_bcpo} b) and \ref{pic_spinladder_dm}. 
As a result we obtain relations between the different bonds and therefore relations
between the components of the $\mathbf{D}$-vectors. 
For a better understanding, we exemplarily demonstrate the different steps of 
the symmetry analysis for the vector $\mathbf{D}_{1}$ corresponding to the NN bonds in detail in App.\ \ref{chap_symmetry_D1}.

This symmetry analysis can be carried out for the vectors 
$\mathbf{D}_{0}$ and $\mathbf{D}_{2}$ as well, see Apps.\ \ref{chap_symmetry_D0}
and \ref{chap_symmetry_D2} for a detailed explanation. 
At this point we just give the results, see Table \RM{1}.

\begin{table}[htb]
\begin{center}
\begin{tabular}{c|c|c}
$D_{ij}^{\alpha}$ & along the legs & parity\\
\hline
$D_{0}^{y}$ & alternating & odd\\
$D_{1}^{x}$ & uniform & odd\\
$D_{1}^{y}$ & alternating & odd\\
$D_{2}^{x}$ & uniform & odd\\
$D_{2}^{z}$ & alternating & even
\end{tabular}
\caption{Behavior of the sign along the legs of the spin ladder and 
the parity with respect to the symmetry SS$_{xy}$ 
of the $\mathbf{D}$ vectors. Components not listed 
vanish due to symmetry arguments. The parity of $D_{0}^{y}$ does not refer to the component itself, but to the corresponding term in the Hamiltonian.}
\label{symmetriesD}
\end{center}
\end{table}

\subsection{Symmetries of the symmetric $\mathbf{\Gamma}$-components}

The components $\Gamma_{ij}^{\alpha\beta}$ of the tensor $\Gamma_{ij}$ represent
 the symmetric anisotropic exchange between the two spin 
components $S_{i}^{\alpha}$ and $S_{j}^{\beta}$. We choose the tensor $\Gamma_{ij}$ 
to be traceless because any finite trace can be incorporated
in the isotropic interaction $J\mathbf{S}_{i}\mathbf{S}_{j}$. 
Furthermore, the tensor has to be symmetric. We derive the formula for the 
components $\Gamma_{ij}^{\alpha\beta}$ based on the $\mathbf{D}$ vectors below.

According to Shekhtman \textit{et al.} \cite{shekh92} it is possible to map two 
coupled spins
\begin{equation}
\label{hamiltonian_rotated}
\mathcal{H}=J\mathbf{S}_{1}\mathbf{S}_{2}+\mathbf{D}\left(\mathbf{S}_{1}\times\mathbf{S}_{2}\right)+\mathbf{S}_{1}\Gamma\mathbf{S}_{2}
\end{equation}
with antisymmetric and symmetric anisotropic interactions onto an isotropic model in a rotated basis. The reason is that the anisotropic interactions are induced by SOC which
results in a rotation of the spin in the hopping from site 1 to site 2.

To keep the calculations transparent we consider two interacting 
spins $\mathbf{S}_{1}$ and $\mathbf{S}_{2}$. The isotropic coupling is denoted with $J$ and the antisymmetric and  symmetric anisotropic interaction by the vector $\mathbf{D}$ and by 
 the tensor $\Gamma$, respectively. More precisely, Shekhtman \textit{et al.} state that the 
Hamiltonian in \eqref{hamiltonian_rotated} is equivalent to the Hamiltonian 
\begin{equation}
\label{hamiltonian_not_rotated}
\mathcal{H}=J^{\prime}\mathbf{S}_{1}\mathbf{S}_{2}^{\prime}
\end{equation}
where $\mathbf{S}_{2}^{\prime}$ is a rotated spin.

For the renormalized isotropic coupling $J'$ between the two spins the relation $J^{\prime}=\frac{4|t|^{2}}{U}$ holds in leading order as it is well-known from the derivation of
the Heisenberg coupling from a Hubbard model \cite{harri67}. 
The hopping amplitude is given by $t$ and the $U$ denotes the repulsion 
energy between two spins on one site. 
Without loss of generality, we choose the $z$-axis as the rotation axis for 
$\mathbf{S}_{2}^{\prime}$. At the end of this 
calculation we will generalize the direction of the rotation axis. Therefore the relation between the spins $\mathbf{S}_{2}^{\prime}$ and $\mathbf{S}_{2}$ 
is given by
\begin{equation}
\label{rotatedS2}
\mathbf{S}_{2}^{\prime}=\begin{pmatrix}
                         \cos\left(\varphi\right) & \sin\left(\varphi\right) & 0\\
			  -\sin\left(\varphi\right) & \cos\left(\varphi\right) & 0\\
			  0 & 0 & 1
                        \end{pmatrix}\mathbf{S}_{2},
\end{equation}
where $\varphi$ is the angle of rotation which is of the order of the SOC.
Using \eqref{rotatedS2} we transform \eqref{hamiltonian_not_rotated} to 
\begin{equation}
\mathcal{H}=J\mathbf{S}_{1}\mathbf{S}_{2}+
J\left(\sqrt{1+\frac{\mathbf{D}^{2}}{J^{2}}}-1\right)S_{1}^{z}S_{2}^{z}+
\mathbf{D}\left(\mathbf{S}_{1}\times\mathbf{S}_{2}\right)
\end{equation}
with the substitutions $J=J^{\prime}\cos\left(\varphi\right)$ and 
$\mathbf{D}=J^{\prime}\sin\left(\varphi\right)\mathbf{e}_{z}$. 
It is reasonable to assume that the 
absolute value of the vector $\mathbf{D}$ is much smaller than the isotropic coupling $J$. 
Thus, we can expand the term 
$\sqrt{1+\frac{\mathbf{D}^{2}}{J^{2}}}=1+\frac{\mathbf{D}^{2}}{2J^{2}}$ 
in leading order. 

Now we generalize the calculation, this means that 
$\mathbf{D}$ points into an arbitrary direction. Then, the Hamiltonian takes the form 
\begin{equation}
\label{hamiltonian_transformed}
 \mathcal{H}=J\mathbf{S}_{1}\mathbf{S}_{2}+
\frac{\mathbf{D}^{2}}{2J^{2}}S_{1}^{\mathbf{D}}S_{2}^{\mathbf{D}}+
\mathbf{D}\left(\mathbf{S}_{1}\times\mathbf{S}_{2}\right).
\end{equation}
The components $S_{i}^{\mathbf{D}}$ represent the component of the spin $\mathbf{S}_{i}$ pointing in $\mathbf{D}$-direction. It is given by 
the projection $S_{i}^{\mathbf{D}}=\frac{\mathbf{D}\mathbf{S}_{i}}{\sqrt{\mathbf{D}^{2}}}$. The antisymmetric part has already 
the correct form, cf.\ \eqref{hamiltonian_rotated}. 
We write down the other two terms component by component to reach a formula for 
the entries $\Gamma^{\alpha\beta}$ depending on the components of $\mathbf{D}$ and the isotropic coupling $J$. Splitting \eqref{hamiltonian_transformed} into its components we obtain
\begin{equation}
\mathcal{H}=\sum_{\alpha,\beta}S_{1}^{\alpha}\left(J\delta^{\alpha\beta}+
\frac{D^{\alpha}D^{\beta}}{2J}\right)S_{2}^{\beta}+
\mathbf{D}\left(\mathbf{S}_{1}\times\mathbf{S}_{2}\right).
\end{equation}
Keeping in mind that the trace of $\Gamma^{\alpha\beta}$ has to vanish we write
\begin{equation}
\mathcal{H}=\sum_{\alpha\beta}S_{1}^{\alpha}\left(\tilde{J}\delta^{\alpha\beta}+\Gamma^{\alpha\beta}\right)S_{2}^{\beta}
+\mathbf{D}\left(\mathbf{S}_{1}\times\mathbf{S}_{2}\right)
\end{equation}
by using the substitutions
\begin{subequations}
\begin{align}
\tilde{J}&=J+\frac{\mathbf{D}^{2}}{6J}\\
\Gamma^{\alpha\beta}&=\frac{D^{\alpha}D^{\beta}}{2J}-
\frac{\delta^{\alpha\beta}\mathbf{D}^{2}}{6J}.
\end{align}
\end{subequations}

We  emphasize that the isotropic coupling is now given by $\tilde{J}$ and not by $J$. 
But due to the assumption that the absolute value of 
$\mathbf{D}$ is much smaller than $J$ the approximation $\tilde{J}\approx J$ is justified.
The deviation is of second order in $D$ (or $\varphi$) only. 
Therefore, the general formula for the entries 
of the tensor $\Gamma_{ij}$ is given by
\begin{equation}
\label{formular_gamma}
\Gamma_{ij}^{\alpha\beta}=\frac{D_{ij}^{\alpha}D_{ij}^{\beta}}{2J_{ij}}-
\frac{\delta^{\alpha\beta}\mathbf{D}_{ij}^{2}}{6J_{ij}}.
\end{equation}
At this point we stress once more that all isotropic interactions are shifted
to the isotropic coupling $\tilde{J}$. 
The tensor $\Gamma_{ij}$ contains only anisotropic interactions and thus has trace zero, 
see \eqref{formular_gamma}. 
In the literature, also other representations of $\Gamma_{ij}$ are in use \cite{plumb14,plumb15}
without vanishing trace.

On the basis of \eqref{formular_gamma}, we translate the properties of the
DM vectors in Table \RM{1} to properties of the matrix elements of the
symmetric tensor $\Gamma$ in Table \RM{2}. This concludes the section on
the general properties of the anisotropic couplings.

\begin{table}[htb]
\begin{center}
\begin{tabular}{c|c|c}
$\Gamma_{ij}^{\alpha\beta}$ & along the legs & parity\\
\hline
$\Gamma_{0}^{xx}$ & uniform & even\\
$\Gamma_{0}^{yy}$ & uniform & even\\
$\Gamma_{0}^{zz}$ & uniform & even\\
\hline
$\Gamma_{1}^{xx}$ & uniform & even\\
$\Gamma_{1}^{xy}$ & alternating & even\\
$\Gamma_{1}^{yy}$ & uniform & even\\
$\Gamma_{1}^{zz}$ & uniform & even\\
\hline
$\Gamma_{2}^{xx}$ & uniform & even\\
$\Gamma_{2}^{xz}$ & alternating & odd\\
$\Gamma_{2}^{yy}$ & uniform & even\\
$\Gamma_{2}^{zz}$ & uniform & even
\end{tabular}
\caption{Behavior of the sign along the legs of the spin ladder and the parity with respect to the symmetry SS$_{xy}$ 
of the components $\Gamma_{ij}^{\alpha\beta}$. Components not listed 
vanish due to \eqref{formular_gamma} or they are given
 by their equivalent expression $\Gamma_{ij}^{\beta\alpha}$.}
\label{symmetriesGamma}
\end{center}
\end{table}


\section{Method}
\label{method}

Here we provide details how we calculate the dispersion in presence of 
the DM interactions and the symmetric anisotropic exchanges.
As described in Sect.\ \ref{chap_isotropic}, the results of the isotropic spin ladder with 
an interladder coupling $J^{\prime}$ are our starting point. Their calculation is performed
by a deepCUT using the 1n-generator up to order 13 in $x$.

In the present article, we focus on bilinear terms stemming from the anisotropic interaction terms because they are the only ones influencing the dispersion on the mean-field level.
Thus, we treat the DM interactions by a mean-field approach justified by the smallness
of the effect. Recall that the interladder coupling is dealt with on the same level.
More sophisticated treatments are subject of future research.

\subsection{Derivation of the bilinear DM terms}
\label{chap_transform_hamiltonian}

We proceed as follows:
\begin{itemize}
\item[1.]{Write down the anisotropic interaction term in the basis of the spin operators 
$S_{i}^{\alpha,\mathrm{L}/\mathrm{R}}$.}
\item[2.]{On the linear operator level the deepCUT maps the spin operators 
$S_{i}^{\alpha,\mathrm{L}/\mathrm{R}}$ onto the effective spin operators 
as in Eqs.\ \eqref{effective_spinoperator}, \eqref{eq:leftright}, and \eqref{eq:effspin}.}
\item[3.]{We treat the triplon operators as bosonic operators in a mean-field approach and 
apply a Fourier transformation.}
\end{itemize}
After these steps we obtain the effective anisotropic interaction terms in momentum space $k$.

We illustrate these steps for the component $D_{2}^{z}$.
It is the only one with even parity, see Table \ref{symmetriesD}, 
which implies that no other component contributes on the bilinear level
due to the odd parity of the triplon creation and annihilation operators.
We emphasize, however, that the other $\mathbf{D}$-components may and will have
contributions on the level of odd numbers of triplon operators.
This means that they may generate linear or trilinear contributions.
Their treatment is beyond the scope of the present article and left to future research.

First, we write down the corresponding anisotropic interaction term
\begin{equation}
\label{HamiltonianD2z}
\mathcal{H}_{\mathrm{NNN},z}^{\mathrm{D}}=
\sum_{i}\sum_{\tau\in\{\mathrm{L,R}\}}D_{2_{i}}^{z,\tau}
\left(\mathbf{S}_{i}^{\tau}\times\mathbf{S}_{i+2}^{\tau}\right)_{z}.
\end{equation}
The index $\tau$ indicates the left (L) and the right (R) leg of the spin ladder, the 
index $i$ stands for the rung. The component $D_{2}^{z}$ 
has even parity and an alternating sign, see Table \ref{symmetriesD}, which means that 
$D_{2_{i}}^{z,\mathrm{L}}=D_{2_{i}}^{z,\mathrm{R}}=D_{2}^{z}\left(-1\right)^{i}$ holds. 
In Step 2 we replace the spin operators in 
\eqref{HamiltonianD2z} by the effective spin operators \eqref{effective_spinoperator}
which yields the  effective anisotropic interaction
\begin{equation}
\mathcal{H}_{\mathrm{NNN},z}^{\mathrm{D,eff}}=2D_{2}^{z}\sum_{i}\left(-1\right)^{i}
S_{i,\mathrm{eff}}^{x,\mathrm{L}}\left(S_{i+2,\mathrm{eff}}^{y,\mathrm{L}}-
S_{i-2,\mathrm{eff}}^{y,\mathrm{L}}\right).
\end{equation}
Expressing this term in bosonic operators and performing a Fourier 
transformation leads to
\begin{align}
\nonumber
\mathcal{H}_{\mathrm{NNN},z}^{\mathrm{D,eff}}&=
4D_{2}^{z}\mathrm{i}\sum_{k}a\left(k\right)a\left(k+\pi\right)\sin\left(2k\right)
\\
\label{HamiltonianD2zeff}
&\phantom{=}\left(t_{k}^{x,\dagger}\left(t_{-k-\pi}^{y,\dagger}+
t_{k+\pi}^{y}\right)-\mathrm{h.c.}\right).
\end{align}
Now we see that the component $D_{2}^{z}$ couples the $x$-mode with momentum $k$ 
to the $y$-mode with momentum $k+\pi$  and the $y$-mode with momentum $k$ to 
the $x$-mode with momentum $k+\pi$.

\subsection{Computation of the dispersion}
\label{computingdispersion}

On the level of bilinear triplon operators treated as standard bosons we 
have to find the appropriate generalized Bogoliubov transformation in order
to diagonalize the Hamiltonian. At present, we only need the dispersion, i.e.,
the eigen energies, without constructing the full diagonalizing
transformation. 
The eigen energies are the eigen values of finite matrices
which we determine in the following way.

We consider the commutator
\begin{equation}
\label{commutator}
 \left[\mathcal{H},v\right]=w
\end{equation}
with the operators $v$ and $w$ which are linear combinations of bosonic operators 
$\mathcal{B}_i$ with  prefactors  $v_{i}$ and $w_{i}$.
The operator structure of $v$ and $w$ is identical, only the prefactors differ. 
The commutation with $\mathcal{H}$ in \eqref{commutator} 
provides linear relations between the prefactors $v_{i}$ and $w_{i}$
which can be cast into the matrix-vector product
\begin{equation}
\label{eq:matrix}
\mathcal{M}\vec{v}=\vec{w}
\end{equation}
where 
\begin{equation}
[\mathcal{H},\mathcal{B}_i]= \sum_j M_{ij} B_j.
\end{equation}
Then we are looking for the eigen values $\lambda$ fulfilling
\begin{equation}
\label{eq:eigenvalue}
\mathcal{M}\vec{v}=\lambda \vec{v}
\end{equation}
for the eigen vector $\vec{v}$.

Thus, we diagonalize the matrix $\mathcal{M}$. The positive eigen values $\lambda$ 
depending on the momentum $k$ represent the dispersion of the considered Hamiltonian 
$\mathcal{H}$.  To find the matrix $\mathcal{M}$ it is useful to identify
 a minimal closed ansatz for the operators $\mathcal{B}_i$. 
The closure means that the commutation with $\mathcal{H}$ of the set $\{ \mathcal{B}_i \}$
does not yield operators which cannot be expressed by $\{ \mathcal{B}_i \}$.
The set should be minimal for convenience because a small number of operators
requires a matrix with low dimension only.
Generally, our ansatz comprises the adjoint operators as well, i.e., if
$\mathcal{B}_i$ is element of our set of operators then $\mathcal{B}_i^\dag$ as well.
This implies that all eigen values come in pairs of positive and negative values.
The positive values result from the creation of a diagonal boson as in 
$[\omega b^\dag b,b^\dag]=\omega b^\dag$ while the negative ones correspond to the
annihilation of a diagonal boson as in
$[\omega b^\dag b,b]= -\omega b$.

For the effective Hamiltonian of the single frustrated spin ladder 
 \eqref{hamiltonian_isotropic_solved} the minimal closed ansatz for $v$ is simply given by 
\begin{equation}
 v_{\text{ladder}}=v_{1}t_{k}^{\alpha,\dagger},
\end{equation}
containing only one operator. 
The corresponding matrix $\mathcal{M}_{\text{ladder}}$ has just one entry which is
\begin{equation}
\mathcal{M}_{\text{ladder}}=\omega_{0}\left(k\right)
\end{equation}
defining the dispersion. Note that this is an exceptional case because
no adjoint operators are considered.

Next, we consider $\mathcal{H}_{\mathrm{NNN},z}^{\mathrm{D,eff}}$ in 
\eqref{HamiltonianD2zeff} where a minimal closed 
ansatz for $v$ is given by
\begin{equation}
\label{ansatz_v}
v=v_{1}t_{k}^{x,\dagger}+v_{2}t_{k+\pi}^{y,\dagger}+v_{3}t_{-k}^{x}+v_{4}t_{-k-\pi}^{y}
\end{equation}
leading to the commutation matrix
\begin{equation}
\mathcal{M}_{D_{2}^{z}}=\begin{pmatrix}
                                0 & \mathrm{i}D_{2}^{z}\left(k\right) & 0 & -\mathrm{i}D_{2}^{z}\left(k\right)\\
				-\mathrm{i}D_{2}^{z}\left(k\right) & 0 & \mathrm{i}D_{2}^{z}\left(k\right) & 0\\
				0 & \mathrm{i}D_{2}^{z}\left(k\right) & 0 & -\mathrm{i}D_{2}^{z}\left(k\right)\\
				-\mathrm{i}D_{2}^{z}\left(k\right) & 0 & \mathrm{i}D_{2}^{z}\left(k\right) & 0\\
                               \end{pmatrix},
\end{equation}
using
\begin{equation}
\label{D2zk}
 D_{2}^{z}\left(k\right)=4D_{2}^{z}a\left(k\right)a\left(k+\pi\right)\sin\left(2k\right).
\end{equation}

Following this pattern, we set up matrices for the isotropic effective Hamiltonian of the 
single spin ladder, all DM interactions, and the interladder coupling. 
Then we diagonalize their sum to obtain the wanted dispersion
from the momentum dependent positive eigen values.

Up to this point, we analyzed the DM interactions and found that only one 
component, $D_{2}^{z}$, contributes to the dispersion. Although the symmetric anisotropic exchanges 
are of second order in SOC we know that they can be equally important \cite{shekh92}. 
To include the symmetric anisotropic exchanges we repeat the steps from Sect.\ 
\ref{chap_transform_hamiltonian} to transform the corresponding 
observables, see App.\ \ref{transformed_anisotropic_terms}.  As discussed before
only components $\Gamma_{ij}^{\alpha\beta}$ of even parity 
contribute to the bilinear Hamiltonian. Finally, the corresponding commutation
matrix $\mathcal{M}$ is computed and added to the other matrices.
We find that the coupling between the $x$-mode and $y$-mode
is modified while the $z$-mode is still separated.

The sum of all matrices for the $x$- and $y$-mode has the form
\begin{equation}
\mathcal{M}_{\mathrm{all}}=\begin{pmatrix}
                  A_{\omega} & -\mathrm{i}B & -A\left(k\right) & \mathrm{i}B\\
		  \mathrm{i}B & C_{\omega} & \mathrm{i}B & -C\\
		  A\left(k\right) & -\mathrm{i}B & -A_{\omega} & \mathrm{i}B\\
		  \mathrm{i}B & C & -\mathrm{i}B & C_{\omega}
                  \end{pmatrix},
\end{equation}
where the entries depend on momentum $k$. Here we used the shorthands
\begin{subequations} 
\begin{align}
\label{Aomega}
A_{\omega}&\coloneqq\omega_{1}+A\left(k\right)\\
C_{\omega}&\coloneqq\omega_{2}+C\\
\label{Ak}
A\left(k\right)&\coloneqq d_{1}+\Gamma_{0}^{xx}\left(k\right)+
\Gamma_{1}^{xx}\left(k\right)+\Gamma_{2}^{xx}\left(k\right)\\
B&\coloneqq\Gamma_{1}^{xy}\left(k\right)-D_{2}^{z}\left(k\right)\\
\label{Ck}
C&\coloneqq d_{2}+\Gamma_{0}^{yy}\left(k\right)+\Gamma_{1}^{yy}
\left(k\right)+\Gamma_{2}^{yy}\left(k\right).
\end{align}
\end{subequations}
The abbreviations in  \eqref{Ak} to \eqref{Ck} stand for
\begin{subequations}
\begin{align}
\label{omega1}
\omega_{1}&=\omega_{0}\left(k\right)\\
\label{omega2}
\omega_{2}&=\omega_{0}\left(k+\pi\right)\\
\label{d1}
d_{1}&=-2J^{\prime}\cos\left(2\pi l\right)a^2\left(k\right)\\
\label{d2}
d_{2}&=-2J^{\prime}\cos\left(2\pi l\right)a^2\left(k+\pi\right)\\
\Gamma_{0}^{xx}\left(k\right)&=-2\Gamma_{0}^{xx}a^2\left(k\right)\\
\Gamma_{1}^{xx}\left(k\right)&=4\Gamma_{1}^{xx}a^2\left(k\right)\cos\left(k\right)\\
\Gamma_{2}^{xx}\left(k\right)&=4\Gamma_{2}^{xx}a^2\left(k\right)\cos\left(2k\right)\\
\label{Gamma1xyk}
\Gamma_{1}^{xy}\left(k\right)&=4\Gamma_{1}^{xy}a\left(k\right)a\left(k+\pi\right)\sin\left(k\right)\\
\label{Gamma0yyk}
\Gamma_{0}^{yy}\left(k\right)&=-2\Gamma_{0}^{yy}a^2\left(k+\pi\right)\\
\label{Gamma1yyk}
\Gamma_{1}^{yy}\left(k\right)&=-4\Gamma_{1}^{yy}a^2\left(k+\pi\right)\cos\left(k\right)\\
\label{Gamma2yyk}
\Gamma_{2}^{yy}\left(k\right)&=4\Gamma_{2}^{yy}a^2\left(k+\pi\right)\cos\left(2k\right).
\end{align}
\end{subequations}

The resulting eigen values read 
\begin{equation}
\label{omega_x}
\omega_{x}\left(k\right) = \sqrt{\frac{1}{2}\Omega_{1}^{2}\pm\frac{1}{2}
\sqrt{\Omega_{2}^{2}+16\omega_{1}\omega_{2}B^{2}}}
\end{equation}
with
\begin{subequations}
\begin{align}
\Omega_{1}&\coloneqq\omega_{1}^{2}+2\omega_{1}A\left(k\right)+\omega_{2}^{2}+2\omega_{2}C\\
\Omega_{2}&\coloneqq\omega_{1}^{2}+2\omega_{1}A\left(k\right)-\omega_{2}^{2}-2\omega_{2}C.
\end{align}
\end{subequations}
One finds that the dispersion of the $y$-mode can be found from the
dispersion of the $x$-mode by a shift by $\pi$
\begin{equation}
\omega_{y}\left(k\right)=\omega_{x}\left(k+\pi\right).
\end{equation}

The analysis of the $z$-mode reveals that it is not coupled to the $x$- and the $y$-mode
at all. Only the symmetric anisotropic exchange has an effect on the $z$-mode. The minimal
closed set only requires two operators for $v$ 
\begin{equation}
v_{z}=v_{1,z}t_{k}^{z,\dagger}+v_{2,z}t_{-k}^{z}.
\end{equation}
The sum of the commutation matrices affecting 
the $z$-mode has the form
\begin{equation}
\mathcal{M}_{\mathrm{all},z}=\begin{pmatrix}
                              \omega_{1}+D & -D\\
			      D & -\omega_{1}-D
                             \end{pmatrix}
\end{equation}
with the abbreviations
\begin{subequations}
\label{abbreviations_z}
\begin{align}
D&\coloneqq d_{1}+\Gamma_{0}^{zz}\left(k\right)+\Gamma_{1}^{zz}\left(k\right)+\Gamma_{2}^{zz}\left(k\right)\\
\Gamma_{0}^{zz}\left(k\right)&\coloneqq-2\Gamma_{0}^{zz}a^2\left(k\right)\\
\Gamma_{1}^{zz}\left(k\right)&\coloneqq4\Gamma_{1}^{zz}a^2\left(k\right)\cos\left(k\right)\\
\Gamma_{2}^{zz}\left(k\right)&\coloneqq4\Gamma_{2}^{zz}a^2\left(k\right)\cos\left(2k\right).
\end{align}
\end{subequations}
The positive eigen values of the matrix $\mathcal{M}_{\mathrm{all},z}$ read
\begin{equation}
\omega_{z}\left(k\right) = \sqrt{\omega_{1}^{2}+2\omega_{1}D}.
\end{equation}


\section{Discussion of the results}

Prior to any attempt to fit the experimental
dispersion by adjusting the anisotropic couplings
we studied the effects of each $\mathbf{D}$-component on $\omega_{x}\left(k\right)$
 separately. We summarize the results in Table \ref{effectsD}.

\begin{table}[htb]
\begin{tabular}{ c | c | c }
$D_{ij}^{\alpha}$ & lin. & effect on $\omega_{x}\left(k\right)$\\
\hline
$D_{0}^{y}$ & \ding{55} & increase in the complete Brillouin zone\\
$D_{1}^{x}$ & \ding{55} & asymmetric shift about $k=\frac{\pi}{2}$\\
& & $\rightarrow$ lowering at $k>\frac{\pi}{2}$\\
$D_{1}^{y}$ & \ding{55} & asymmetric shift about $k=\frac{\pi}{2}$\\
& & $\rightarrow$ lowering at $k<\frac{\pi}{2}$\\
$D_{2}^{x}$ & \ding{55} & lowering around the minimum \\
$D_{2}^{z}$ & \ding{51} & linear effect: shift minimum to higher $k$-values\\
& & quadratic effect: increase around the minimum
\end{tabular}
\caption{Effects of an increase of the various $\mathbf{D}$-components on the dispersion 
$\omega_{x}\left(k\right)$. If the component induces an effect
in linear order it is marked by \ding{51}, otherwise we put \ding{55}. 
All components contribute in quadratic order, i.e., via the symmetric $\Gamma$-components.}
\label{effectsD}
\end{table}

Based on this understanding of the effects of 
anisotropic couplings we systematically searched for values of the $\mathbf{D}$-components 
which provide the best match between the calculated dispersion and the measured dispersion data. We departed from the isotropic coupling ratios
$x=1.2$ and $y=0.9$ and used the calculated 
isotropic dispersion $\omega_{0}\left(k\right)$ and the coefficients $a_{\delta}$
resulting from the transformation of the observable.
Then, we looked for appropriate values of the $\mathbf{D}$-components and of the energy scale 
$J_{0}$. Below, we indicate the $\mathbf{D}$-components in units 
of the corresponding isotropic coupling, i.e., 
we use $\widetilde{D}_{i}^{\alpha}=\nicefrac{D_{i}^{\alpha}}{J_{i}}$ with $i\in\{0,1,2\}$
labeling the various bonds, see Fig.\ \ref{pic_spinladder_dm}.

\begin{figure}[htb]
\includegraphics[width=\columnwidth]{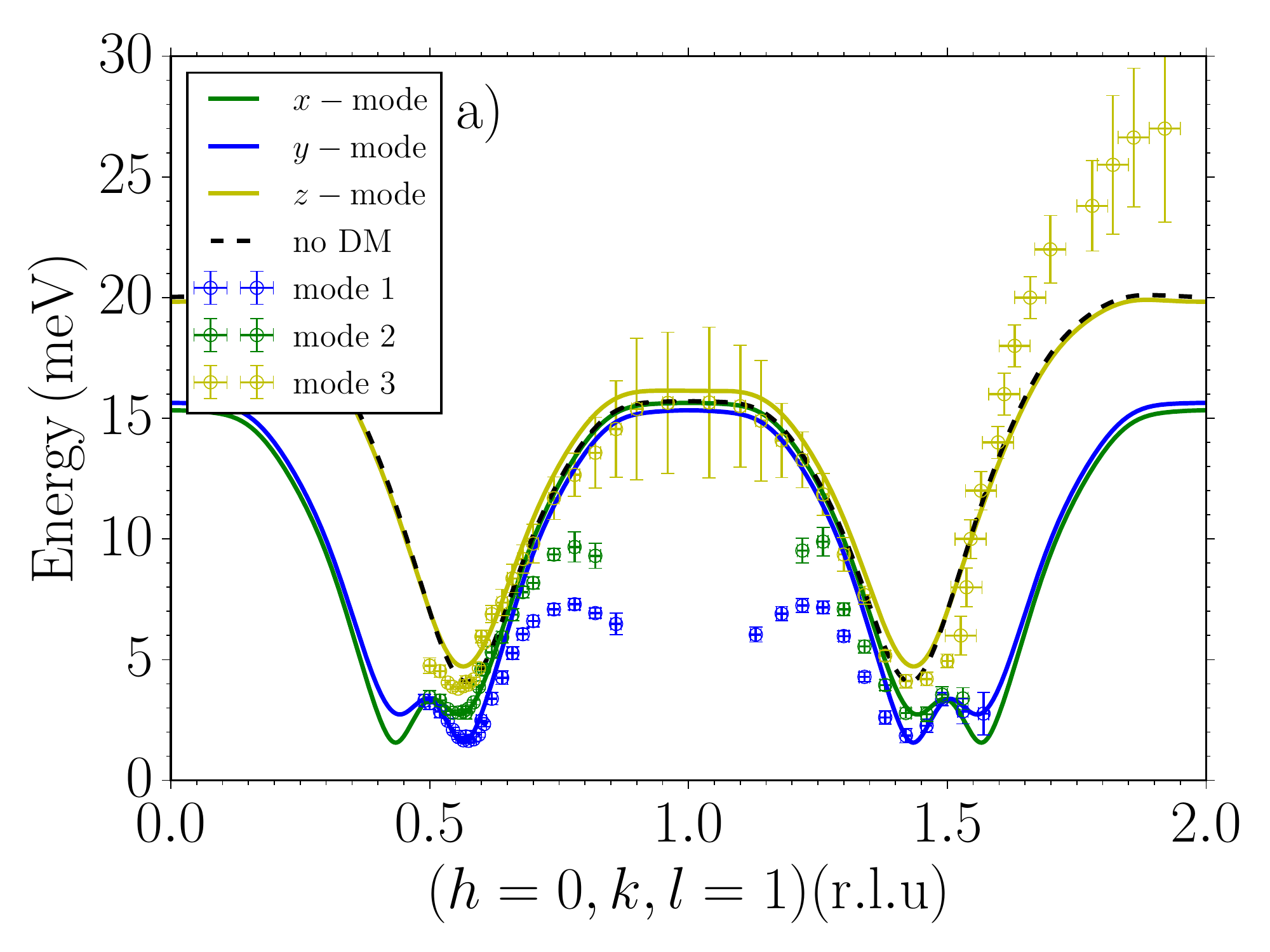}
\includegraphics[width=\columnwidth]{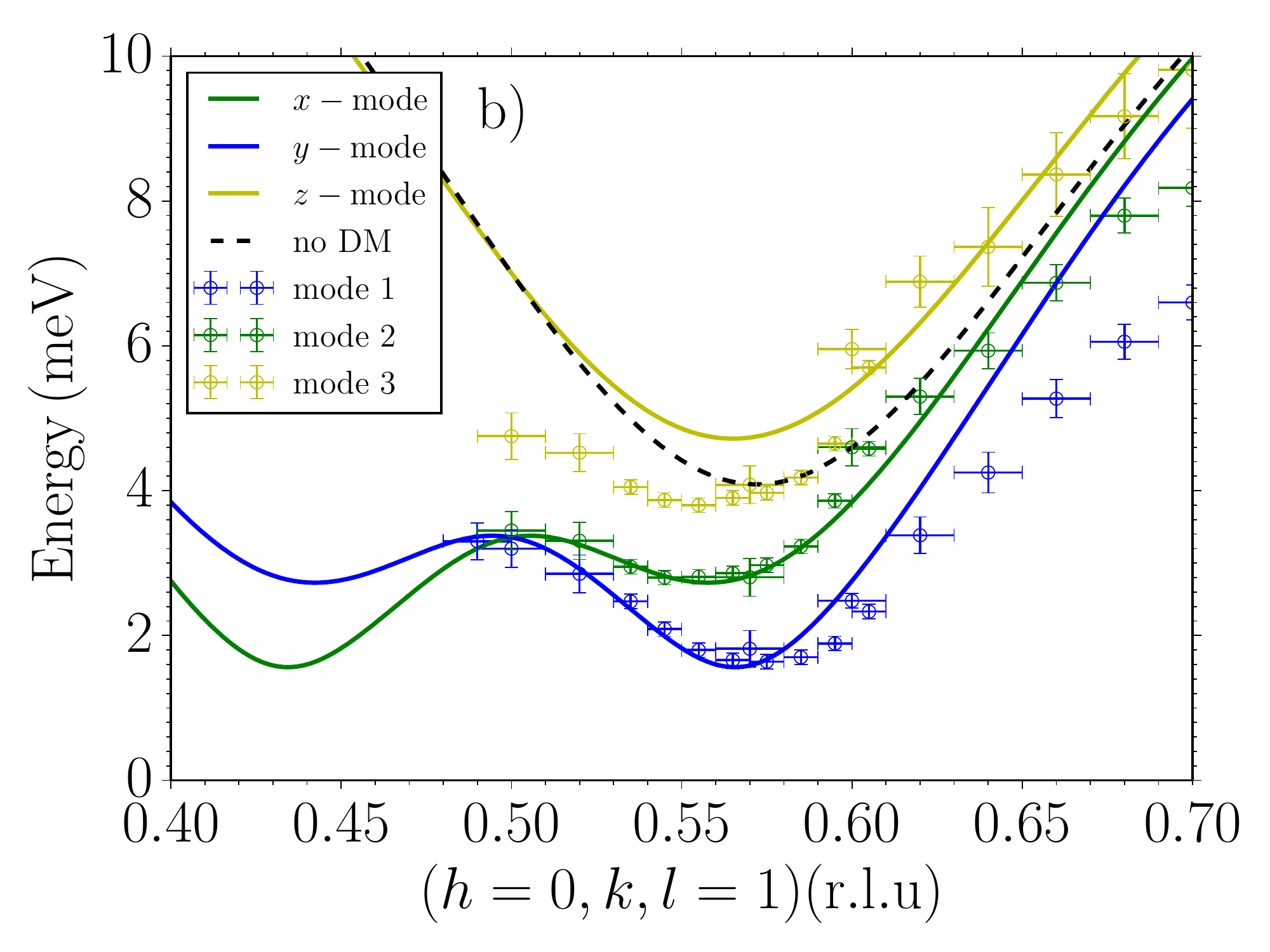}
\caption{a) Fitted theoretical dispersions $\omega_{\alpha}$, $\alpha\in\{x,y,z\}$ 
for $x=1.2$, $y=0.9$ and $J^{\prime}=1.5$\,meV. 
The fitted parameters are $J_{0}=9.4\,\mathrm{meV}$, 
$\widetilde{D}_{0}^{y}=0.00$, $\widetilde{D}_{1}^{x}=0.48$, 
$\widetilde{D}_{1}^{y}=0.61$, $\widetilde{D}_{2}^{x}=0.00$,
 and $\widetilde{D}_{2}^{z}=-0.02$. b) Zoom of panel a) into the vicinity 
of the left minimum.}
\label{x_12_y_09}
\end{figure}

In the following, we discuss several issues concerning the theoretical fits
depicted in Fig.\ \ref{x_12_y_09}.

{\bf (i)} A good description of the measured data in the area of the minimum of mode 1 
and mode 2 is achieved with the calculated dispersions 
$\omega_{x}\left(k\right)$ and $\omega_{y}\left(k\right)$. 
The necessary large values of the components $\widetilde{D}_{1}^{x}$ and 
$\widetilde{D}_{1}^{y}$ represent an unsatisfying feature. We expected the relative anisotropic 
couplings to assume values of $\widetilde{D}\approx 0.1-0.2$. 
The reason why one has to choose such large values for $\widetilde{D}_{1}^{x}$ and 
$\widetilde{D}_{1}^{y}$ is that one needs $\Gamma_{1}^{xy}$ to be sufficiently large. 
We found out that this term leads to the lowering of the dispersions 
$\omega_{x}\left(k\right)$ and $\omega_{y}\left(k\right)$ around the point $k=0.5\,\left(\text{r.l.u}\right)$.\
 and to a flattening of the W-shape of the dispersions. At the point 
$k=0.5\,\left(\text{r.l.u}\right)$. the dispersion without anisotropic interactions takes the value 7.00\,meV, 
the experimental values of mode 1 and 2 take the values 3.20\,meV and 3.55\,meV. 
This implies that the anisotropic interactions 
have to lower the dispersions at $k=0.5$ about 3 to 4\,meV. 
To achieve such a large energy difference the component $\Gamma_{1}^{xy}$, which 
causes the main influence on the dispersion at $k=0.5$, has to accept 
a large value. Hence, the components $\widetilde{D}_{1}^{x}$ and $\widetilde{D}_{1}^{y}$ 
have to assume large values due to  the relation \eqref{formular_gamma}. 
We choose $\widetilde{D}_{1}^{y}$ to be slightly larger than $\widetilde{D}_{1}^{x}$ to create 
the slightly asymmetric behavior of the measured dispersion about $k=0.5\,\left(\text{r.l.u}\right)$. We stress 
that it is possible to swap the dispersions of the 
$x$- and $y$-mode by swapping the values of 
$\widetilde{D}_{1}^{x}$ and $\widetilde{D}_{1}^{y}$.

{\bf (ii)} The chosen value of $\widetilde{D}_{2}^{z}$ is negative and small. 
The reason for the sign of the component 
can be found in Eq. \eqref{omega_x}. Only for a negative sign 
the effects of $\Gamma_{1}^ {xy}$ and 
$\widetilde{D}_{2}^{z}$ partly compensate so that the value of the minimum is 
approximated in a satisfying way.

{\bf (iii)} Major discrepancies between the shape of the calculated dispersion of the 
$z$-mode and the measured mode 3 cannot be eliminated. The measured data shows 
a W-shaped dispersion like for the modes 1 and 2. But the overall shape of the 
calculated $z$-mode is similar to the dispersion \emph{without} anisotropic interactions, 
see Fig.\ \ref{x_12_y_09}, panel a). The only difference between the two curves is that 
the $z$-mode is slightly increased about the minimum 
by finite $\Gamma_{0}^{zz}$, $\Gamma_{1}^{zz}$, and $\Gamma_{2}^{zz}$, see Fig.\
 \ref{x_12_y_09}, panel b).

{\bf (iv)} Around $k=0.75\,\left(\text{r.l.u}\right)$ the two lowest modes bend towards lower energies. 
The corresponding theoretical modes do not show this feature. We expect that
inclusion of the two-triplon continuum and its hybridization with the one-triplon
states will explain this feature, see Refs.\ \onlinecite{fisch11a,fisch11bb,zhito13,zhito06} for similar
calculations of asymmetric spin ladders. The importance
of the two-triplon continua  has already been pointed out by Plumb {\it et al.}
\cite{plumb15}. But so far no theoretical description of the down-bending exists
to our knowledge. We come back to this point in Sect.\ \ref{summary}.

{\bf (v)} The maximum value reached by the $z$-mode is $\approx 19$\,meV. 
The measured maximum value is $\approx 27$\,meV. 
We tried hard to obtain a better match
between experiment and theory at high energies and did  not succeed.
Other ratios $x$ and $y$ do not help in this respect either. 
In view of the large error bars it is reasonable to presume
that states of higher triplon number and the hybridization with them
need to be taken into account. This is beyond the scope of the present
article and subject of future research.

We recall that it was our aim to describe the experimentally measured dispersion 
in BCPO by including anisotropic interactions. We assumed these interactions to accept values between 10\,$\%$ and 20\,$\%$ of the isotropic couplings. 
Summarizing, we state that this was not possible. Large values
of $\mathbf{D}_{1}\approx0.6J_{1}$ must be assumed 
to achieve agreement between  experiment and theory. 
Even then the $z$-mode cannot be described convincingly at low energies.
Moreover, the broad resonances at high energies are not captured
either.

Our results go well with the ones from Plumb \textit{et al.} \cite{plumb14,plumb15}. They chose the couplings constants to assume the following values based on bond operator theory (BOT)
on the mean-field level:
$x=1$, $y=1$, $J_{0}=8\,\mathrm{meV}$, $J^{\prime}=1.6\,\mathrm{meV}$, 
$\widetilde{D}_{1}^{x}=0.6$ and $\widetilde{D}_{1}^{y}=0.4$.
The parameters are in good agreement, i.e., they differ only by up to 
20\%. Thus the comprehensive high-order CUT approach confirms the BOT results and refines
them.  The large values of the 
DM interactions, however, do not fulfill our expectation for 
the anisotropy in the exchange of copper spins as discussed above. 

A striking discrepancy between the experimental and the calculated dispersion is 
the shape of the evaluated $z$-mode.
To improve the shape, it is necessary to identify an interaction which couples the 
$z$-mode with momentum $k$ with the $z$-mode with momentum $k+\pi$. As we have seen in our previous analysis, this type of interaction has the effect that the dispersion splits up into 
an upper and a lower branch yielding a shallow W-shape if the coupling is large enough.
So far, we have not found such a coupling, but we will consider possible candidates 
in the next section.


\section{Alternating  next-nearest neighbor coupling}
\label{sect_alternatingJ2}

Here, we want to discuss possible extension of the model considered so far
which may help to understand and to describe the magnetism in BCPO better.

\begin{figure}[htb]
\centering
\includegraphics[scale=0.12]{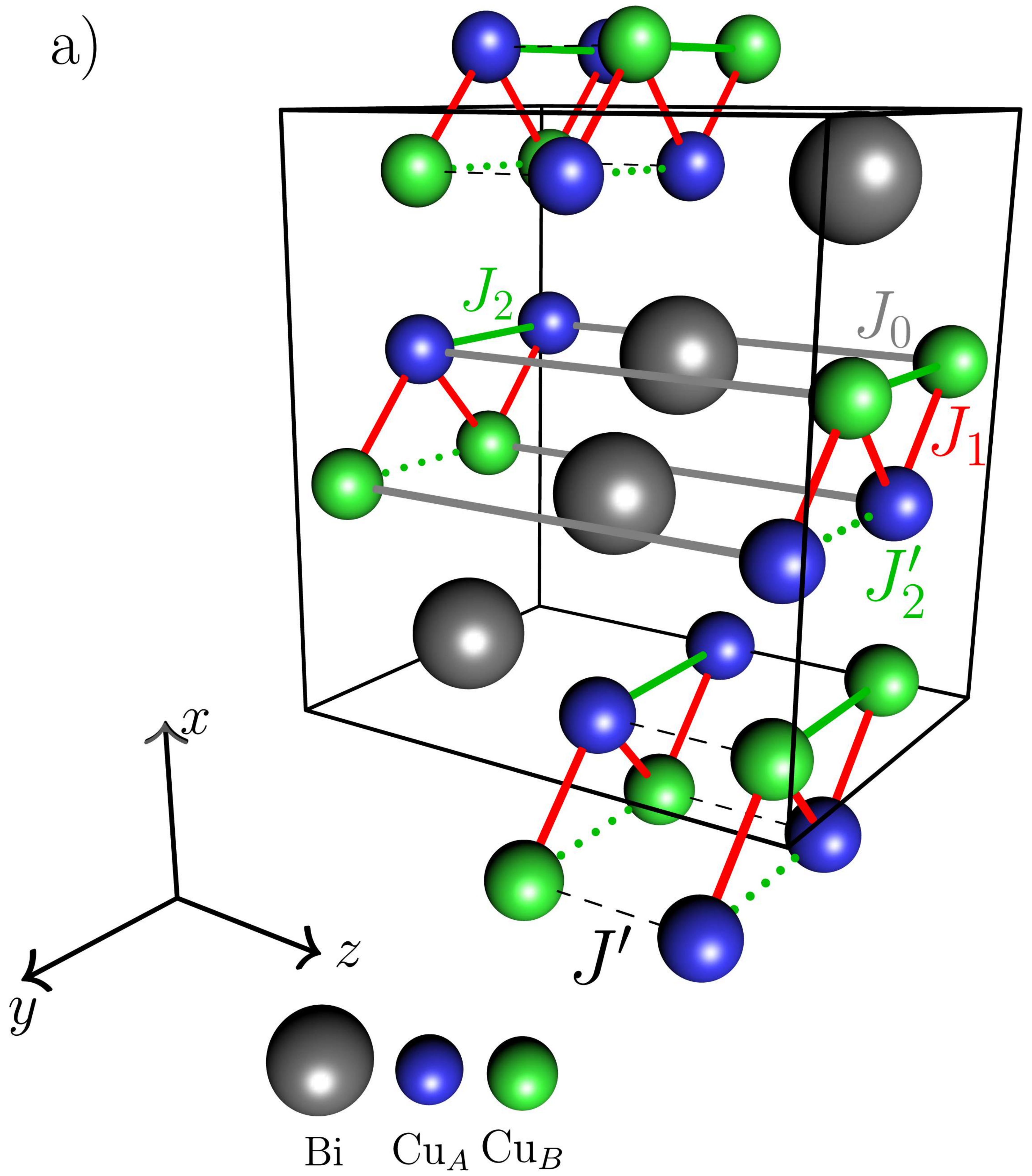}
\includegraphics[scale=0.12]{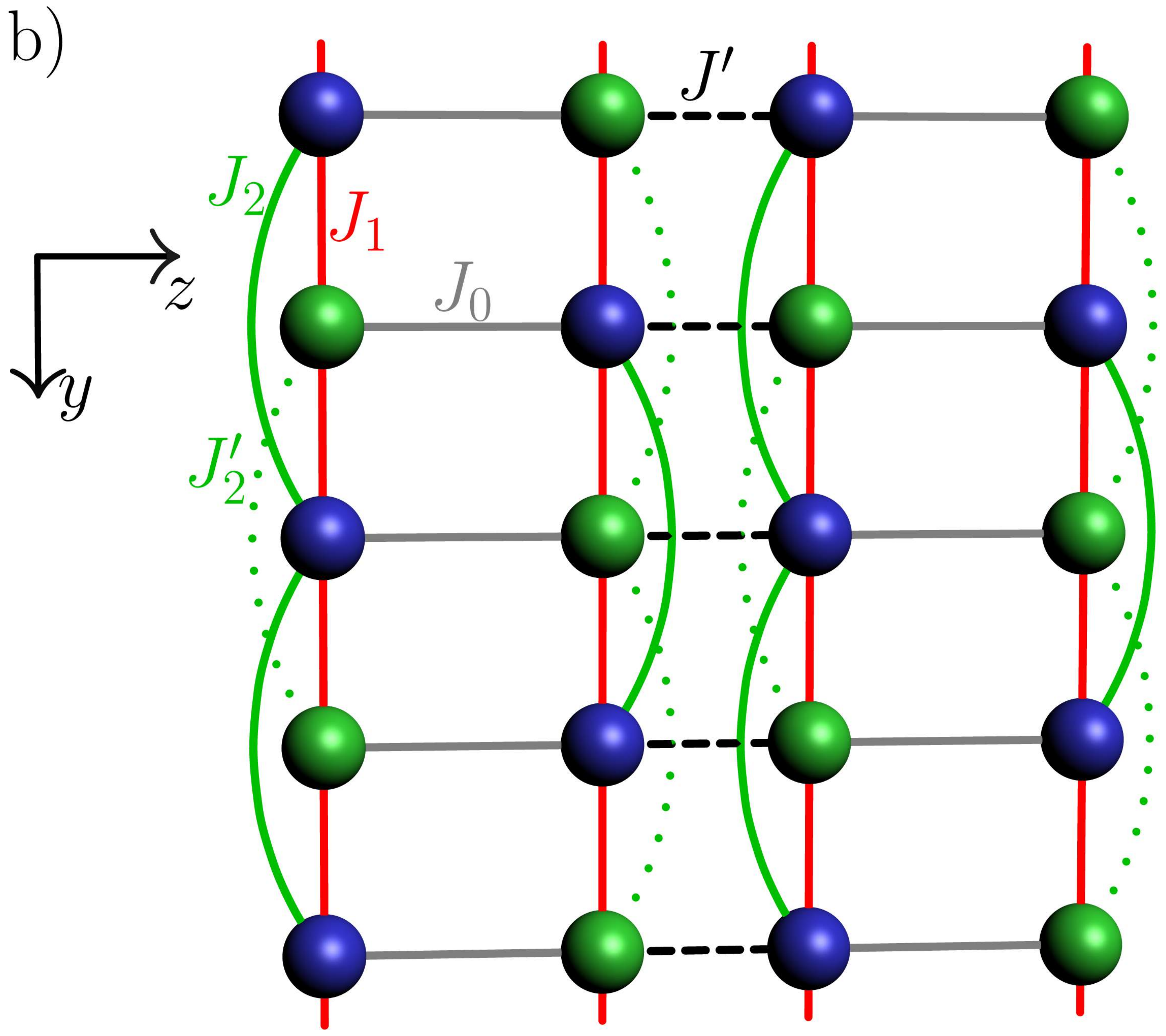}
\includegraphics[scale=0.12]{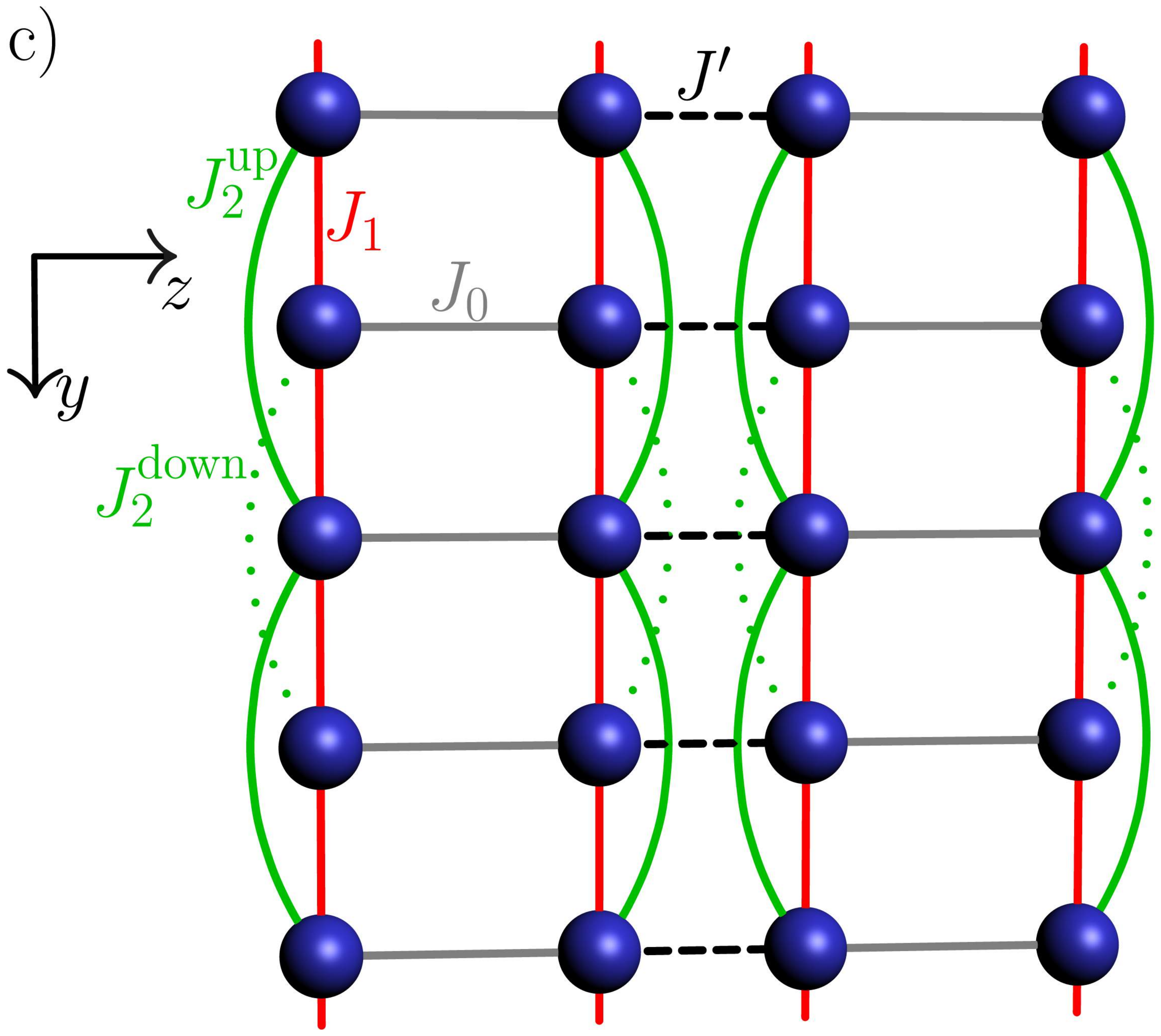}
\caption{a) Crystal structure of BCPO including the alternation of $J_{2}$. The variation of the coupling $J_{2}$ is visualized 
by the two different couplings $J_{2}$ and $J_{2}^{\prime}$. \\
b) Effective spin model analyzed by Tsirlin \textit{et al.} \cite{tsirl10} including the alternation NNN coupling. 
Here, the inequivalence of the copper ions is taken into account and therefore the alternation of the NNN 
coupling has odd parity.\\
c) Effective spin model including the alternation of $J_{2}$. The analyzed model is made of frustrated spin ladders with an 
alternating NNN coupling, which are coupled by an interladder coupling $J^{\prime}$. Again the inquivalence of the 
copper ions is neglected so that the alternation of the NNN coupling has even parity.}
\label{pic_structure_bcpo_alt}
\end{figure}

The first idea suggesting itself is to consider the differing copper ions, see
Fig.\ \ref{pic_structure_bcpo_alt}. The coupling $J_2$ among the Cu$_A$ 
and the coupling $J_2'$ among the Cu$_B$ can be different. Tsirlin 
\textit{et al.} \cite{tsirl10} computed it and found that it is
significantly large. The relative difference can be quantified by 
$r:=(J_2'-J_2)/J_2$. Inspecting Fig.\ \ref{pic_structure_bcpo_alt} b) 
we see that $r$ changes sign by shifting the ladder by one NN bond
along the legs. Of course, this can only be done if we view the
ladder as being flat which we can do for the sake of symmetry analysis.
Thus, this alternation indeed couples modes at $k$ to those at $k+\pi$.

But in addition, $r$ has odd parity, i.e., it changes sign if the
spin ladder is reflected at its center line. This implies that it
will be represented by terms of odd number of triplon operators.
Thus on the level of our description no effect will ensue. 
But even if we computed the effects of these terms in infinite order
of perturbation it would not yield a coupling of the triplon mode
at $k$ to one at $k+\pi$ because due to the \emph{odd} parity of the perturbation
quantified by $r$ this would require an \emph{even} number of application
of the perturbing Hamiltonian. Hence, the overall momentum change would
be an even  multiple of $\pi$ equivalent to zero.
We conclude that this term does not suffice to explain the
observed shallow W-shape of the $z$-mode.

Therefore, we vary the alternation of $J_2$, see Fig.\ \ref{pic_structure_bcpo_alt} c). 
We \emph{assume} that it is even at the
temperatures at which the magnetism is measured. This means, that we assume
that the couplings $J_2$ is the same along the rails of the tubes in Fig.\ 
\ref{pic_structure_bcpo_alt} c) and it is the same in each layer of the tubes. But it
differs between the lower layer and the upper layer by an alternation 
$\delta:=(J_2^{\text{up}}-J_2^{\text{down}})/(J_2^{\text{up}}+J_2^{\text{down}})$.
The key point is that this alternation is \emph{even} with respect to reflections
of the spin ladder about the center line and it is \emph{alternating} 
along the (flattened) spin ladder. Hence, it is capable to couple the modes
at $k$ to the modes at $k+\pi$. This is the empirical reason why we introduce
this kind of alternation. At present, it is not backed by structural analyses
of the crystal at low temperatures to our knowledge. We like to point out that only
small shifts of the order of 1\% in the atomic positions are required to justify
the values we will use for $\delta$, see below, because the magnetic couplings
are extremely sensitive to the precise position values.
We suggest that the low temperature structure is re-analyzed in this respect.

We will show below that the alternation $\delta$ of $J_2$ indeed improves the
fits of the magnetic dispersions at low energies considerably. In contrast,
an alternation of the NN coupling $J_{1}$ has hardly an effect around $k=\pi/2$
because its matrix element contains the factor $\cos\left(k\right)$ in the effective observable. 

\subsection{Inclusion of the alternation in the NNN coupling}

The term in the Hamiltonian representing this alternation reads
\begin{equation}
\label{hamiltonianJ2}
\mathcal{H}_{J_{2}}=J_{2}\delta\sum_{i,\tau}
\left(-1\right)^{i}\mathbf{S}_{i}^{\tau}\mathbf{S}_{i+2}^{\tau}.
\end{equation}
We include this term in a perturbative way. 
As described in Sect.\ \ref{method} the first step is to insert the
effective spin operators \eqref{effective_spinoperator} and to transform 
the resulting expression to $k$ space yielding
\begin{align}
\nonumber
\mathcal{H}_{J_{2}}^{\mathrm{eff}}&=2J_{2}\delta
\sum_{k,\alpha}a\left(k\right)a\left(k+\pi\right)\cos\left(2k\right)\\
\label{HamiltonianJ2eff}
&\phantom{=}\left(t_{k}^{\alpha,\dagger}t_{-k-\pi}^{\alpha,\dagger}+
2t_{k}^{\alpha,\dagger}t_{k+\pi}^{\alpha}+t_{k}^{\alpha}t_{-k-\pi}^{\alpha}\right).
\end{align}
As expected the effective term  $\mathcal{H}_{J_{2}}^{\mathrm{eff}}$ couples 
modes with momentum $k$ and momentum $k+\pi$ of each flavor $\alpha$. 
The alternation $\delta$ is multiplied with $\cos\left(2k\right)$, which means that it gives a 
contribution at $k=0.5\,\left(\text{r.l.u}\right)$ corresponding to $k=\pi/2$ in the theoretical description.

\subsection{Symmetry analysis of the $\mathbf{D}$-components}

The alternation $\delta$ lowers the symmetry of the crystal structure. In concrete terms, this
means that the two symmetries RS$_{y}$ and SS$_{yz}$ of the five symmetries in Sect.\ 
\ref{symmetriesDcomponents} are not fulfilled any more. Therefore, it is necessary to 
perform the complete symmetry analysis again. We present the results of the symmetry 
analysis in Table \ref{symmetriesDalternating}.

\begin{table}[htb]
\begin{center}
\begin{tabular}{c|c|c}
$D_{ij}^{\alpha}$ & along the legs & parity\\
\hline
$D_{0}^{y}$ & - & odd\\
$D_{1}^{x}$ & uniform & odd\\
$D_{1}^{y}$ & alternating & odd\\
$D_{1}^{z}$ & uniform & even\\
$D_{2}^{x}$ & - & odd\\
$D_{2}^{z}$ & - & even
\end{tabular}
\caption{Behavior of  the sign along the legs of the spin ladder with  NNN alternation 
$\delta$ and the parity of the $\mathbf{D}$ vectors. 
Components not listed have to vanish due to symmetry arguments. 
The symbol "-" means that it is not possible to determine the behavior 
of the sign with the help of the present symmetries.}
\label{symmetriesDalternating}
\end{center}
\end{table}

The most interesting result of the symmetry analysis is that the component $D_{1}^{z}$ 
does not have to vanish any more. The parity of this 
component is even which means that $D_{1}^{z}$ 
provides a contribution to the dispersion on bilinear level. 
We presume that the best matching value for $\delta$ ranges between
 10\,$\%$ and 15\,$\%$ because this is roughly the value required to lower the
isotropic dispersion of the uniform spin ladder to the experimental values
around $k=0.5\,\left(\text{r.l.u}\right)$. As a consequence, we assume that the component $D_{1}^{z}$ 
accepts value between 10\,$\%$ and 15\,$\%$ of the components $D_{1}^{x}$ and $D_{1}^{y}$ 
because the finite values of $D_{1}^{z}$ only results from the additional symmetry breaking
by the NNN alternation $\delta$.

The analysis of the $\Gamma$-components shows that the parity of the previously
non-vanishing components does not change. The parity of the components
which do not vanish because of the contribution of $D_{1}^{z}$ have odd parity.
Thus, they do no influence the dispersion. The only effect of 
$D_{1}^{z}$ on the $\Gamma$-components is a certain change of the value of the components 
$\Gamma_{1}^{\alpha\alpha}$ according to \eqref{formular_gamma}. 

For the linear effect of $D_{1}^{z}$ we express the outer product in spin space
\begin{equation}
\mathcal{H}_{\mathrm{NN},z}^{\mathrm{D}}=\sum_{i,\tau}D_{1_{i}}^{z,\tau}
\left(\mathbf{S}_{i}^{\tau}\times\mathbf{S}_{i+1}^{\tau}\right)
\end{equation}
in terms of triplon operators as described before in Sect.\ \ref{chap_transform_hamiltonian} 
leading to
\begin{align}
\nonumber
\label{H_eff_D1z}
\mathcal{H}_{\mathrm{NN},z}^{\mathrm{D,eff}}&=4D_{1}^{z}\mathrm{i}\sum_{k}a^2\left(k\right)\sin\left(k\right)
\\
&\phantom{=}\left(t_{k}^{x,\dagger}
\left(t_{-k}^{y,\dagger}+t_{k}^{y}\right)-\mathrm{h.c.}\right).
\end{align}
The $D_{1}^{z}$-component modifies the coupling between the $x$- and $y$-mode 
as the component $D_{2}^{z}$ does. We emphasize that the $D_{1}^{z}$-component couples 
between the modes with same momenta $k$ while the $D_{2}^{z}$-component connects the momenta $k$ and $k+\pi$.

\subsection{Computation of  the dispersion}

To take the influence on the dispersion 
of the alternation $\delta$ and the ensuing component $D_{1}^{z}$ 
into account we have to extend the set of operators $\{\mathcal{B}_i\}$ used before for $v$ in
\ref{ansatz_v} because the commutators 
$\left[\mathcal{H}_{J_{2}}^{\mathrm{eff}},v\right]$ and 
$\left[\mathcal{H}_{\mathrm{NN},z}^{\mathrm{D,eff}},v\right]$ 
yield operators not contained in the previous ansatz for $v$. 
In presence of the alternation, the minimal and complete ansatz is given by
\begin{align}
\nonumber
v_{J_{2}}&=\phantom{+}v_{1}t_{k}^{x,\dagger}+v_{2}t_{k+\pi}^{x,\dagger}+
v_{3}t_{-k}^{x}+v_{4}t_{-k-\pi}^{x}\\
\label{ansatz_v_J2}
&\phantom{=}+v_{5}t_{k}^{y,\dagger}+v_{6}t_{k+\pi}^{y,\dagger}+
v_{7}t_{-k}^{y}+v_{8}t_{-k-\pi}^{y}.
\end{align}
As explained in Sect.\ \ref{computingdispersion} the next step is to commute the
complete effective Hamiltonian with $v_{J_{2}}$ in order to set up the commutation
matrix arising from \eqref{commutator}. The positive eigen values of this matrix
 $\mathcal{M}_{\mathrm{all},xy,J_{2}}$ represent the dispersion of the $x$- and $y$-mode. 
Since the required  ansatz \eqref{ansatz_v_J2} comprises eight operators the resulting 
matrix  is an 8 $\times$ 8 matrix and cannot be diagonalized analytically, 
see App.\ \ref{concrete_8x8matrix}.  Therefore,  the eigen values have to be computed 
numerically.

Considering the $z$-mode we find that it is still not coupled to the $x$- and $y$-mode.
Thus the minimal and complete ansatz for it to include the effect of the NNN
alternation $\delta$ is given by
\begin{equation}
\label{ansatz_v_z_J2}
v_{z,J_{2}}=v_{1}t_{k}^{z,\dagger}+v_{2}t_{k+\pi}^{z,\dagger}+
v_{3}t_{-k}^{z}+v_{4}t_{-k-\pi}^{z}.
\end{equation}
The resulting commutator matrix from \eqref{commutator} is 4 $\times$ 4 reading
\begin{equation}
\mathcal{M}_{\mathrm{all},z,J_{2}}=\begin{pmatrix}
                                    D_{\omega,1} & J_{2} & -D\left(k\right) & -J_{2}\\
				    J_{2} & D_{\omega,2} & -J_{2} & -D\left(k+\pi\right)\\
				    D\left(k\right) & J_{2} & -D_{\omega,1} & -J_{2}\\
				    J_{2} & D\left(k+\pi\right) & -J_{2} & -D_{\omega,2}
                                   \end{pmatrix}
\end{equation}
with the abbreviations
\begin{subequations}
\begin{align}
D_{\omega,1}&=\omega_{0}\left(k\right)+D\left(k\right)\\
D_{\omega,2}&=\omega_{0}\left(k+\pi\right)+D\left(k+\pi\right)\\
\label{J_2}
J_{2}&=4J_{2}\delta a\left(k\right)a\left(k+\pi\right)\cos\left(2k\right).
\end{align}
\end{subequations}
The concrete form of $D\left(k\right)$ is listed in Eqs.\ \eqref{abbreviations_z}. 
The positive eigen values of $\mathcal{M}_{\mathrm{all},z,J_{2}}$ are the following 
\begin{equation}
\omega_{z}\left(k\right)= \sqrt{\frac{1}{2}\Omega_{3}^{2}\pm\frac{1}{2}\sqrt{\Omega_{4}^{2}+
16\omega_{1}\omega_{2}{J_{2}}^{2}}}
\end{equation}
using the shorthands
\begin{subequations}
\begin{align}
\Omega_{3}&=\omega_{1}^{2}+2\omega_{1}D\left(k\right)+\omega_{2}^{2}+2\omega_{2}D\left(k\right)\\
\Omega_{4}&=\omega_{1}^{2}+2\omega_{1}D\left(k\right)-\omega_{2}^{2}-2\omega_{2}D\left(k\right).
\end{align}
\end{subequations}

\subsection{Discussion of the results}

Again, we search for values of the $\mathbf{D}$-components, which provide the best 
match between the measured data and the evaluated dispersions. We start with the results 
of the isotropic ladder with the parameters  $x=1.2$ and $y=0.9$ and fix the interladder 
coupling $J^{\prime}=1.5$\,meV. The NNN alternation $\delta$ and 
the $\mathbf{D}$-components are varied to obtain the best agreement between experiment and theory. 

\begin{figure}[htb]
\includegraphics[width=\columnwidth]{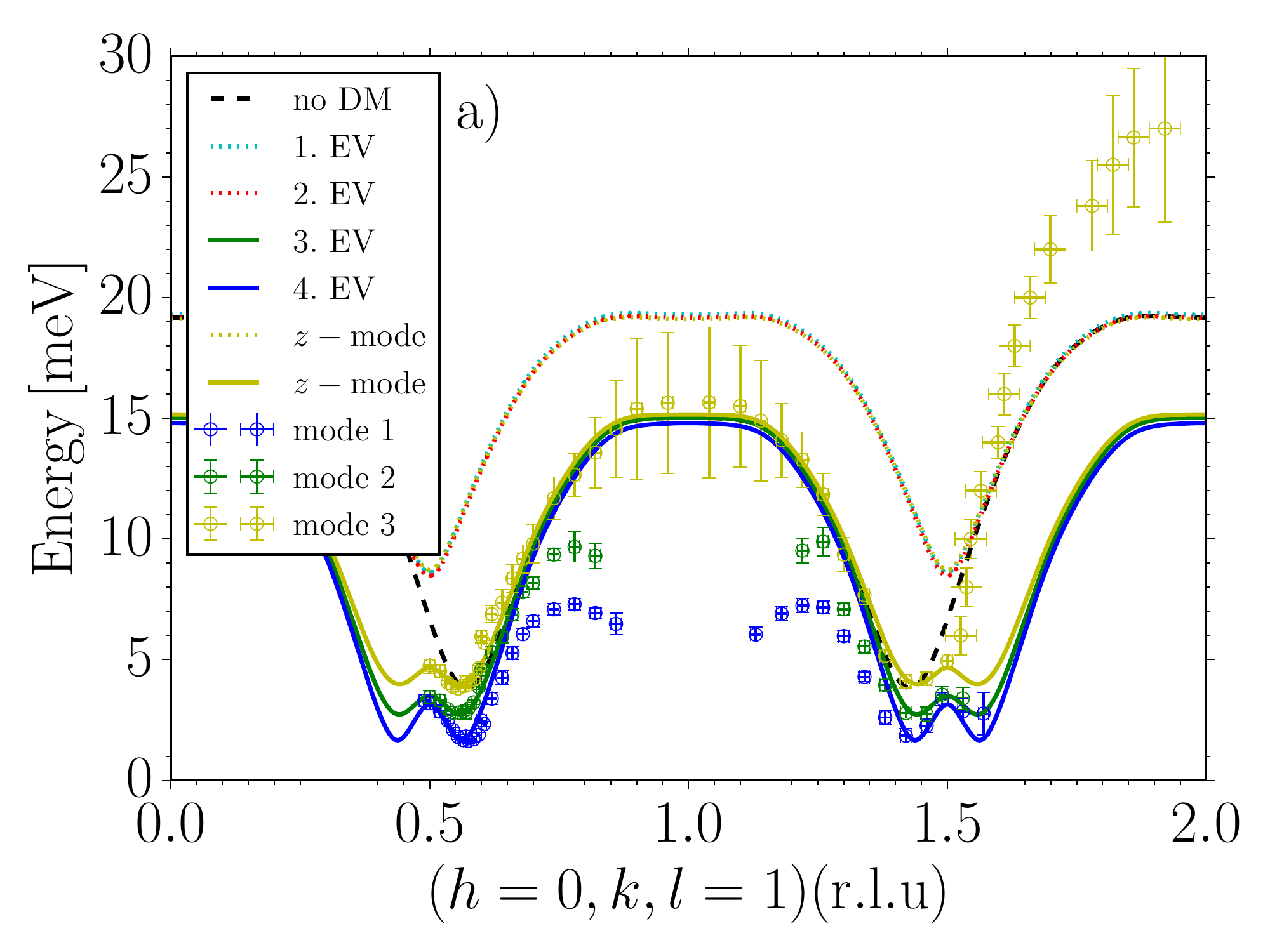}
\includegraphics[width=\columnwidth]{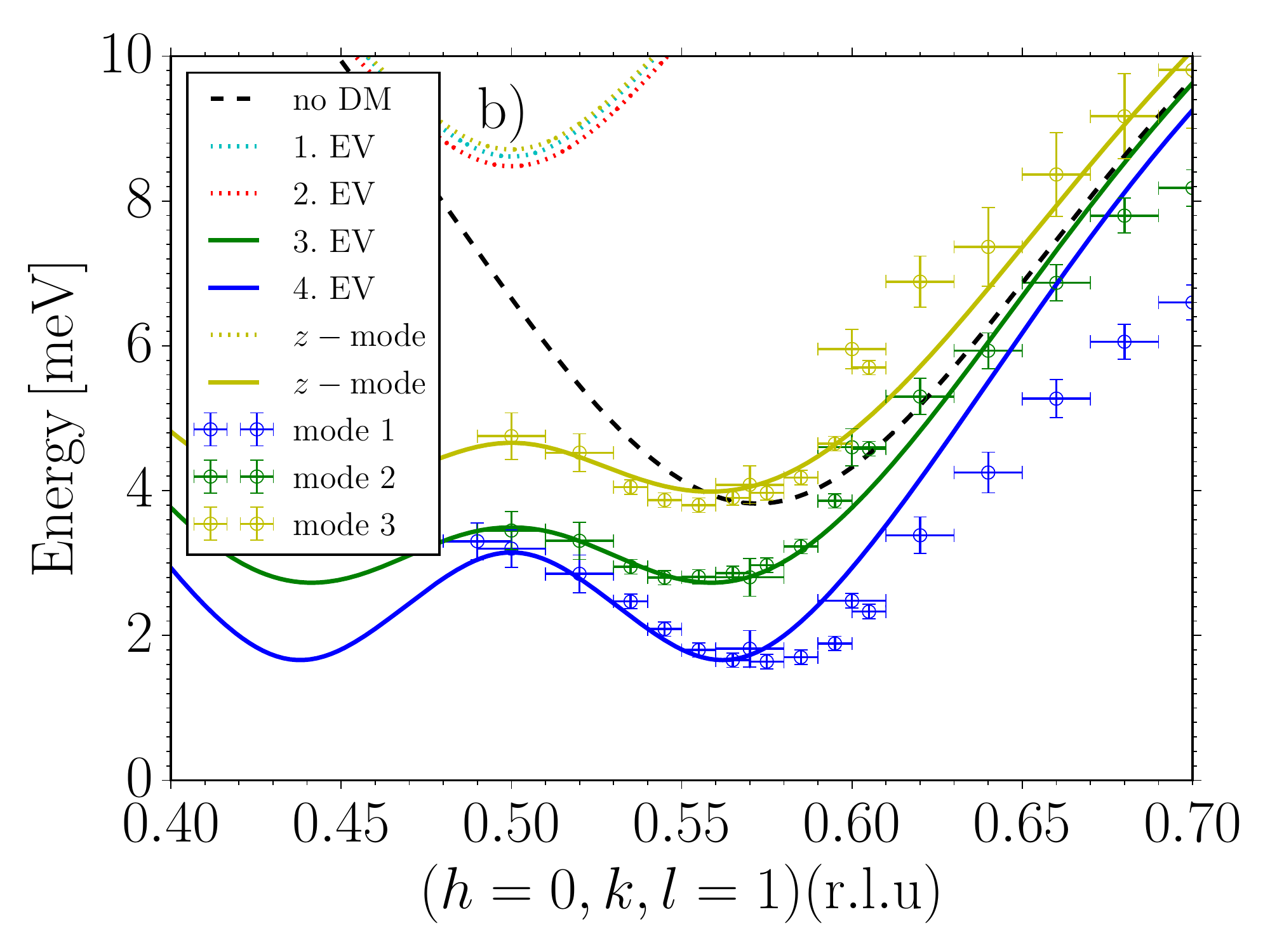}
\caption{a) Fitted  dispersions for fixed values $x=1.2$, $y=0.9$ and $J^{\prime}=1.5$\,meV 
with finite NNN alternation $\delta$. 
The chosen values are: $\delta=0.13$, $J_{0}=9.00\,\mathrm{meV}$, $\widetilde{D}_{0}^{y}=0.35$,
$\widetilde{D}_{1}^{x}=0.36$, $\widetilde{D}_{1}^{y}=0.34$, $\widetilde{D}_{1}^{z}=-0.019$, 
$\widetilde{D}_{2}^{x}=0.28$, and $\widetilde{D}_{2}^{z}=-0.06$. 
The dotted lines are the eigen values of the matrices $\mathcal{M}_{\mathrm{all},xy,J_{2}}$ and 
$\mathcal{M}_{\mathrm{all},z,J_{2}}$ with small weight which do not matter in the
fits, but are shown for the sake of completeness.
b) Enlarged section of panel a) around the left minimum.}
\label{x_12_y_09_J2}
\end{figure}

Below  we discuss several issues of the results depicted in Fig.\ \ref{x_12_y_09_J2}.
 
{\bf (i)} The eigen values $\mathcal{M}_{\mathrm{all},xy}$ provide
four positive energies and the ones of $\mathcal{M}_{\mathrm{all},z}$ two positive energies. 
These six values can be divided into three upper branches and three lower branches 
which have a W-shape.  Fig.\ \ref{x_12_y_09_J2} shows that it is possible to describe the 
three measured modes by the three lowest energies. 
The energies in the upper branch lie clearly above the measured data and are not suitable for a description of the experiment.

{\bf (ii)} We determined the alternation $\delta$ by fitting the evaluated lower 
$z$-dispersion to the measured mode 3 at the $k=0.5\,\left(\text{r.l.u}\right)$.

{\bf (iii)} Comparing the best matching values of $\widetilde{D}_{1}^{x}$ and 
$\widetilde{D}_{1}^{y}$ in presence of the alternation to the values of the previous section we 
clearly see that they can be chosen much lower. Previously, we had to choose the components to accept $\widetilde{D}_{1}^{x}=0.48$ and $\widetilde{D}_{1}^{y}=0.61$. With alternation 
$\delta=0.13$ the values $\widetilde{D}_{1}^{x}=0.36$ and $\widetilde{D}_{1}^{y}=0.34$
are sufficient. As expected the alternation $\delta$ already lowers the dispersion for each
 flavor in the vicinity of $k=0.5\,\text{r.l.u}$. Therefore, $\Gamma_{1}^{xy}$ has 
not the main influence on the $x$- and $y$-dispersion any more and its value
can be  reduced and so the components $\widetilde{D}_{1}^{x}$ and $\widetilde{D}_{1}^{y}$.

Mainly the component $\widetilde{D}_{1}^{z}$ is responsible for the splitting 
at the $k=0.5\,\text{r.l.u.}$ between the two lowest modes resulting from the $x$- and $y$-dispersion. 
In combination with the alternation $\delta$ the components $D_{0}^{y}$ and $D_{2}^{z}$ have a minimal influence 
on the mentioned splitting. 
The value of $\widetilde{D}_{1}^{z}$ is negative and accepts approximately 5$\,\%$ of 
$\widetilde{D}_{1}^{x}$ and $\widetilde{D}_{1}^{y}$ in the reasonable range of values,
see the discussion at the beginning of this section. 

{\bf (v)}  The component $\widetilde{D}_{2}^{z}$ is again chosen negative to achieve a good
match in the vicinity of the minimum.

{\bf (vi)}  The energy values of the minima of the three lowest modes agree nicely 
the theoretical dispersions. Even the value at $k=1$\,r.l.u. matches with the 
measured $z$-mode.

{\bf (vii)} The bending-down behavior of mode 1 and 2 around $k=0.75$\,r.l.u. 
still cannot be described by the modified theory. 
This is another piece of evidence for the necessity to include the hybridization
of the two-particle-continuum in future more extended studies, see for instance
Ref.\ \onlinecite{plumb15}.

{\bf (viii)} The discrepancy at the high energies around $\approx27$\,meV 
persists. The NNN alternation $\delta$ has no important effect on the largest 
evaluated energies.

We also varied the parameters $x$ and $y$ to improve our description of the experimental data, 
but did not reach better results than the ones presented here.

Summarizing this section we are able to describe the three lowest measured dispersions 
with anisotropic interactions of less than $40\,\%$ of the isotropic couplings. 
It was necessary to introduce an alternation $\delta$ in the NNN 
coupling $J_2$ of about $15\,\%$. This alternation lowers the crystal symmetry and as 
a consequence the $D_{1}^{z}$ can be finite producing a small 
splitting between the two lowest modes. 
The qualitative discrepancies between experiment and theory for the two lowest modes around
$k=0.75$\,r.l.u. and at high energies could not be resolved.

The fit could be improved considerably compared to the fit
in the previous section without NNN alternation. First, the dispersion of the
third mode with its shallow W-shape is captured. Second, the values of the relative
DM couplings are significantly closer to reasonable expectations for 
the super exchange between copper ions.
Note in this context that recently, Plumb {\it et al.} \cite{plumb15} also advocated
much smaller values $\tilde D\approx 0.3$ for BCPO when compared to
theoretical calculations including many-triplon states.
This supports our second fit presented in this section with its lower values
for the DM interactions.


\section{Perpendicular Dispersion}

For the sake of completeness we also discuss the dispersion perpendicular to 
the spin ladder, that means in spatial $z$-direction, see Fig.\ \ref{pic_spinladder_dm}. 
The width of its cosine shape is mainly affected by the interladder coupling 
$J^{\prime}$. 

\begin{figure}[htb]
\includegraphics[width=\columnwidth]{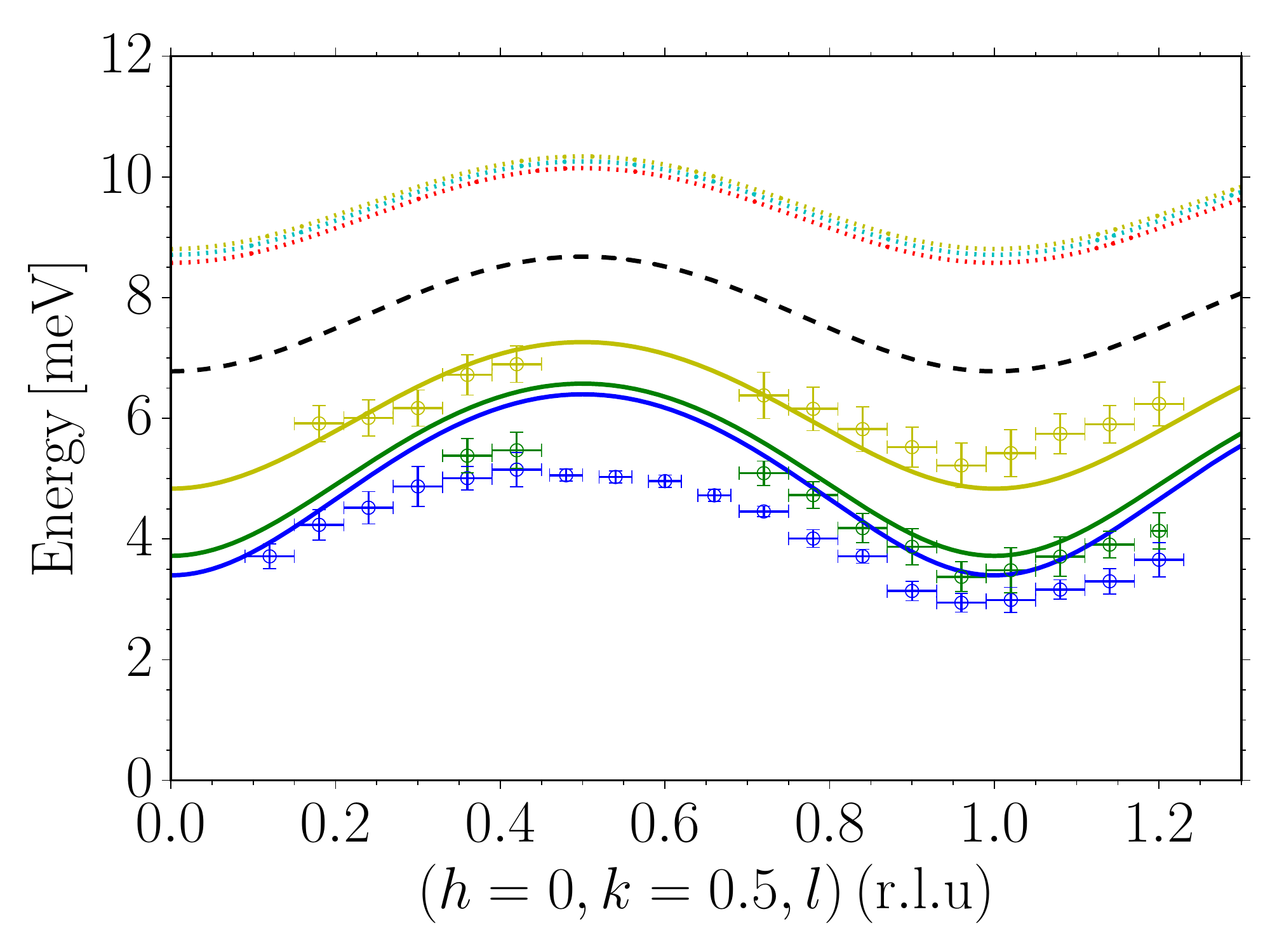}
\caption{Fitted dispersions perpendicular to the spin ladder. The color coding and 
the values of the parameters are the same as in Fig.\ \ref{x_12_y_09_J2}.}
\label{x_12_y_09_J2_perpendicular}
\end{figure}

The shape of the $z$-mode dispersion agrees well with the measured dispersion, see
Fig.\ \ref{x_12_y_09_J2_perpendicular}.
The agreement for the two lower modes is poorer, but still acceptable.
Since a change of $J^\prime$ essentially influences all three modes in the same 
way, it is not possible to reduce the band width of the two lower modes
without affecting the upper $z$-mode. We attempted to find better overall fits
by varying $J^\prime$. Although it is possible to improve the agreement
of the perpendicular dispersion this leads to poorer agreement in the
dispersions along the spin ladders. Thus we still favor the parameter set
used. Fig.\ \ref{x_12_y_09_J2_perpendicular} shows that it yields a reasonable
agreement. 

Moreover, we have to stress again that a description on the 
single-triplon level, i.e., with a bilinear Hamiltonian, cannot capture
all details of BCPO where a significant influence of higher triplon states
is obvious, see the large error bars of the peaks at high energies in
Fig.\ \ref{measured_data}. In particular the two low-lying modes seem 
to hybridize with two-triplon continua as conjectured in Ref.\ \onlinecite{plumb15}
from the down-bending of their dispersions around $k=0.75$\,r.l.u..


\section{Magnetic field dependence}

A model on the single-triplon level can address the effect of magnetic fields as well.
So we turn to this issue as a final check for the validity of the minimal model
advocated. The magnetic field dependence of the three gap values of BCPO has been 
analyzed in the past on experimentally and theoretically 
\cite{koham12,masud03,tsirl10,hwang16a,mentr09,kotes10}.

A magnetic field is incorporated in the Hamiltonian by the Zeeman term 
\begin{equation}
 \mathcal{H}_{Z}=-g\mu_{B}\mathbf{H}\sum_{i,\tau}\mathbf{S}^{\tau}_{i},
\label{eq:zeeman1}
\end{equation}
with the $g$-factor, the Bohr magneton $\mu_{B}$ and the magnetic field $\mathbf{H}$. 
For copper ions as in BCPO, the expected value for  $g$-factor is $2$
or slightly larger by up to 20\%. 
Analyses of the magnetic susceptibility indicate that the $g$-factor of BCPO takes
the value $g\approx 2.1$ due to the 
influence of the bismuth ions \cite{tsirl10,mentr09,kotes10}. \\
We stress that the expected range of $g$-values \cite{Henke} for Cu$^{2+}$ ions is $g=2.1-2.3$ which implies 
at maximum a 15$\,\%$ effect of the SOC. 
Another aspect to be mentioned is the fact that BCPO may contains strong 
magnetoelastic couplings due to its structure, like the copper mineral azurite \cite{crawf}. 
Thus, an easy interpretation of susceptibility measurements is difficult.

Because of the importance of anisotropic interactions in BCPO the $g$-factor 
may be anisotropic as well. This means that it may have different values depending 
on the direction of the magnetic field $\mathbf{H}$. This has be kept in mind in
the following analysis.

Another interesting aspect is that the $g$-tensor may contain a staggered part due to the 
two inequivalent copper ions in a unit cell. To take this effect into account the following arguments 
concerning the transformation of the Zeeman term would not be valid and the calculations 
become significantly more complicated. 
As the total experimental information on the field dependence is presently still limited, 
only a few points are available, see Fig.\ \ref{pic_magnetic_field}, we 
do not discuss this aspect in the present article.

\subsection{Transformation of the Zeeman term}

To proceed we have to identify the transformation of the Zeeman term \eqref{eq:zeeman1}
in terms of triplon operators. At first glance, one may think that the effective spin 
operator \eqref{effective_spinoperator} solves this issue as before. But in fact
the problem is more complicated and simpler at the same time. First, it is
more complicated because the Zeeman term is even with respect to reflection about
the center line. This implies that there is no linear contribution but one has
to pass to the bilinear terms which we have not considered so far.

Second, however, it is simpler because the total spin $\sum_{i,\tau}\mathbf{S}^{\tau}_{i}$
is the generator of global rotations in spin space. Since the CUT is performed for
the isotropic spin ladder conserving spin rotation invariance the total spin is
not altered at all by the CUT. Thus we can compute its representation in terms
of triplon operators prior to any CUT and still use it for the effective model
afterwards. 

Using the general representation of the spin operators by triplon operators \cite{sachd90} 
\begin{align}
S^{\alpha,\mathrm{L/R}}_{i}&=\phantom{-}\frac{1}{2}\Big(\pm t_{i}^{\alpha,\dagger}\pm
t_{i}^{\alpha}- \sum_{\beta,\gamma}
\mathrm{i}\epsilon_{\alpha\beta\gamma}t_{i}^{\beta,\dagger}t_{i}^{\gamma}\Big)
\end{align}
and performing the Fourier transform one obtains straightforwardly
\begin{equation}
\label{magnetic_field_z_momentum}
\mathcal{H}_{Z}=-g\mu_{B}H^{z}\mathrm{i}\sum_{k}\left(t_{k}^{y,\dagger}t_{k}^{x}-
t_{k}^{x,\dagger}t_{k}^{y}\right),
\end{equation}
for a magnetic field in $z$-direction. For magnetic fields in $x$- or $y$-direction 
\eqref{magnetic_field_z_momentum} the spin components only need to be permuted cyclically.

A magnetic field parallel to the $z$-axis as in \eqref{magnetic_field_z_momentum}
induces a coupling between the $x$-mode and the $y$-mode without changing the momentum.
If the magnetic field points into $y$-direction, 
$\mathbf{H}=H^{y}\mathbf{e}_{y}$, a coupling between the $x$-mode and the $z$-mode 
ensues and if the magnetic field has only a $x$-component, $\mathbf{H}=H^{x}\mathbf{e}_{x}$
there is a coupling between the  $y$-mode and the $z$-mode without change of momentum.

We emphasize that the Zeeman term is transformed to a bilinear triplon expression without
any approximation.

\subsection{Computation of the dispersion}

To assess the effect of the magnetic field on  the dispersions 
$\omega_{\alpha}\left(k\right)$, $\alpha\in\{x,y,z\}$ we follow the steps explained in Sect.\
 \ref{computingdispersion}. To this end, we have to find a minimal closed 
set of operators for the ansatz of the input operator $v$. 
In the case of $\mathbf{H}=H^{z}\mathbf{e}_{z}$, the ansatz \eqref{ansatz_v_J2} 
for the coupled $x$- and $y$-mode and the ansatz \eqref{ansatz_v_z_J2} for the 
$z$-mode continue to be appropriate. The reason is that the magnetic field introduces no 
new couplings in addition to the considered anisotropic couplings.

However, for $\mathbf{H}=H^{x}\mathbf{e}_{x}$ or $\mathbf{H}=H^{y}\mathbf{e}_{y}$,
respectively, one has to combine the ansatz  $v_{J_{2}}$ in \eqref{ansatz_v_J2}
 and the ansatz $v_{z,J_{2}}$ in \eqref{ansatz_v_z_J2} leading to 
\begin{align}
\nonumber
v_{\mathrm{mag}}&=\phantom{+}v_{1}t_{k}^{x,\dagger}+v_{2}t_{k+\pi}^{x,\dagger}+v_{3}t_{-k}^{x}+
v_{4}t_{-k-\pi}^{x}\\
\nonumber
&\phantom{=}+v_{5}t_{k}^{y,\dagger}+v_{6}t_{k+\pi}^{y,\dagger}+v_{7}t_{-k}^{y}+
v_{8}t_{-k-\pi}^{y}\\
&\phantom{=}+v_{9}t_{k}^{z,\dagger}+v_{10}t_{k+\pi}^{z,\dagger}+v_{11}t_{-k}^{z}+
v_{12}t_{-k-\pi}^{z}.
\label{eq:alloperators}
\end{align}
No ansatz with less operators is closed under the commutation with the full Hamiltonian.

As explained above in Sect.\ \ref{computingdispersion} one has to set up the commutation
matrix $\mathcal{M}_{\mathrm{all},\mathrm{mag}}$ for the complete Hamiltonian.
Due to the twelve operators in \eqref{eq:alloperators} the matrix is of dimension twelve.
So its eigen values providing the dispersions cannot be computed analytically, but 
the numerical solution is effortless. 

\subsection{Discussion of the results}

\begin{figure}
\includegraphics[width=\columnwidth]{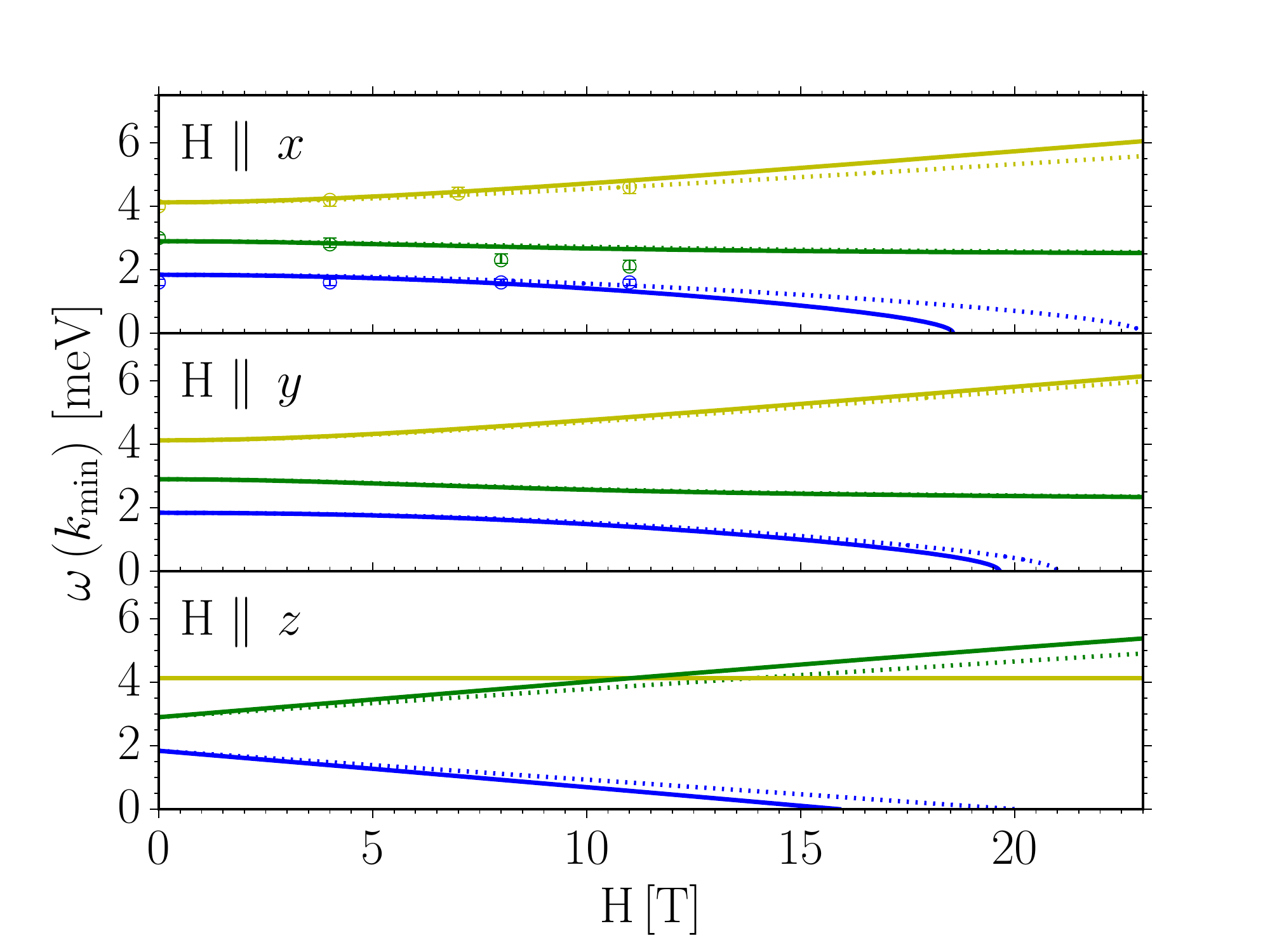}
\caption{Computed gaps of the three lower modes at fixed momentum 
$k_{\mathrm{min}}=0.575$\,r.l.u. for the best fitting parameters as used in 
Fig.\ \ref{x_12_y_09_J2}.
The color coding is the same as in the previous figures. 
The solid lines show the evolution of the three gaps depending on magnetic fields
along the three crystallographic axes for $g=2$. In the upper panel ($H\parallel x$)
we compare our results with the neutron scattering data from Ref.\ \onlinecite{plumb15}. 
The dashed lines result from  anisotropic values $g_{x}=1.6$, $g_{y}=1.9$, and $g_{z}=1.6$
fitted such that the critical fields $H^{x}_{c}=23\,$T, $H^{y}_{c}=21\,$T, and 
$H^{z}_{c}=20\,$T measured by Kohama \textit{et al.} \cite{koham12}
are reproduced.}
\label{pic_magnetic_field}
\end{figure}

Fig.\ \ref{pic_magnetic_field} displays the results for the gap values 
at the incommensurate momentum $k_{\mathrm{min}}=0.575$\,r.l.u..
The solid curves are evaluated with $g=2$.
For magnetic field along the $x$-axis we can compare to experimental data,
see upper panel in Fig.\ \ref{pic_magnetic_field}.
The agreement is very good for the lower and the upper mode. 
The middle mode is reasonably approximated.

The other panels display the effect of magnetic fields
along other directions. In all three directions a critical field $H_c$ exists 
at which the lowest gap closes and
the system enters another phase which can be viewed as a 
condensate of the gapless triplons \cite{fisch11a}.
In comparison to the measured critical fields \cite{koham12}
the theoretical values are too low by about up to 20\%,
see fitted $g$-values given in the caption of Fig.\ \ref{pic_magnetic_field}.

We think that the reason of this discrepancy is the neglect
of the hardcore constraint of the triplons. We know from
the transverse Ising model in one dimension which can be described
either by non-interacting fermions or by hardcore bosons 
that the disordered quantum phase appears to be too unstable
if the bosons are treated as standard bosons. So we conclude that
the closure of the gaps is not quantitatively captured by 
our mean-field type approach. Another aspect is the possible alternation of the g-tensors which we have neglected. 
In view of these arguments
the achieved agreement for the behavior under applied magnetic field
can be considered satisfying.

\section{Summary}
\label{summary}

\subsection{Conclusions}

In this article, we analyzed the influence of anisotropic interactions in the frustrated spin ladder system BiCu$_2$PO$_6$ (BCPO). We presented a single-triplon 
description of the excitation spectrum. The first step was to identify
a starting point for the perturbative treatment of the anisotropic 
couplings. For this we used an advanced version of the continuous unitary transformation,
here deepCUT \cite{krull12}, and computed the dispersion of a single frustrated spin ladder
in a reliable and systematically controlled fashion.

Additionally we included the interladder coupling $J^{\prime}$ on the level
of a mean-field theory. We fixed the interladder coupling in 
units of the rung coupling to $J^{\prime}/J_{0}=0.16$ which we justified
afterwards. We determined the fit parameters $x=J_1/J_0$ and $y=J_2/J_1$
such that the position of the gap in momentum space and the ratio between the dispersion at 
$k=1$\,r.l.u. and the gap value $\Delta$ are described as well as possible.  The best matching values were found to be 
$x\approx 1.2$ and $y\approx 0.9$. Yet the single-triplon
mode at high energies does not match the measured ones which, however,
are very broad suggesting that many-triplon states are needed to
reach a good description.

In a next step, we determined the directions of the $\mathbf{D}$-vectors allowed by symmetry.
As a result of the symmetry analysis of the crystal structure of BCPO, we found that five components out of the nine possible ones may have finite values.
But only one of them has even parity with respect to reflection of the spin ladder
about the center line. In a single-triplon theory only the bilinear terms matter
which are even in parity. Terms with odd number of triplons are odd.
This single $\mathbf{D}$-component is not sufficient to describe BCPO and hence we
extended our analysis also to the symmetric anisotropic $\Gamma$ couplings.
According to Shekhtman {\it et al.} \cite{shekh92}, the symmetric terms
are as important as the antisymmetric ones.

We showed that the $x$- and $y$-mode are coupled by the 
the full set of DM-interactions. The $z$-mode remains uncoupled and can be treated 
separately. The comparison of the computed dispersions with the experimental data 
demonstrated that the two lower measured modes can be described 
well by the theoretical coupled $x$- and $y$-mode 
in the low energy part of the spectrum. But this 
is only possible by assuming unreasonably large anisotropic interactions 
${D}_{1}^{y}\approx 0.6J_1$. This issue also occurred in bond-operator
analyses on mean-field level which starts from coupled dimers 
\cite{plumb14,plumb15}. Another discrepancy is the shape of the computed 
$z$-dispersion $\omega_{z}\left(k\right)$ which does not match to any measured dispersion
such as the upper mode.

In order to improve the description, we conjectured that
BCPO at low temperatures displays an alternation of the next-nearest neighbor coupling $J_2$
which is of even parity, but alternating along the ladders.
Its relative strength is expressed by $\delta$. At this point, 
we stress that this conjectured alternation $\delta$ is not (yet) 
confirmed by structural analysis. The explicit calculation shows
that a value of $\delta=0.13$ leads indeed to a considerably
improved description of the upper mode. Hence, all three modes
are nicely captured at lower energies by our minimal model.
In addition, with this alternation the required values for the 
$D$-components can be lowered to less than $0.6J_1$. 
We expect that the inclusion of many-triplon effects will reduce the required DM 
coupling strengths even further well below $D\approx0.3J$ as indicated by 
diagrammatic perturbation theory \cite{plumb15,hwang16a}. 

For completeness, we analyzed the dispersion perpendicular to the spin ladder as well.
The obtained results agree reasonably well with the measured data.

Finally, we studied the magnetic field dependence of all three energy gaps
and all crystallographic directions for the magnetic field.
For magnetic fields along the $x$-direction experimental data is available
and the critical fields along all three directions.
Our theory describes the finite energy gaps for magnetic fields
along $x$ very well; only the middle mode does not fit perfectly.
The critical fields are reproduced within 20\%. On the one hand, this
is reassuring because it shows that the theory captures the physics
correctly. On the other hand, an even better agreement
would be desirable. We think that the
quantitative discrepancy  is due to the neglect of the hardcore constraint
in our approach.

In total, the present study provides a comprehensive derivation of
a minimal model for the triplon excitations in BCPO  on the single-triplon
level. This means that the effective Hamiltonian is expressed by bilinear
terms of the triplon operators.
The approach is based on a systematically controlled continuous unitary transformation
of the frustrated spin ladder. All other couplings such as interladder couplings,
anisotropic couplings and alternations are included perturbatively on a mean-field 
level. The three low-lying modes are described very well.

\subsection{Outlook}

Our results call for a re-analysis of the crystal structure of BCPO 
at low temperatures. The conjectured alternation of the NNN couplings
along the spin ladder translates to a difference in the couplings
in the upper and in the lower plane of the tube, see Fig.\ \ref{pic_structure_bcpo}.
The improvement of the minimal model including this alternation is
significant and important so that is necessary to verify or to falsify
this point experimentally.

Within the minimal model established above a next theoretical step for improved
understanding is to address the spectral weights quantitatively. To this end,
one would have to compute the eigen vectors of the commutation matrices $\mathcal{M}$
in order to evaluate the overlap of the spin operators occurring in the dynamic 
structure factor with the eigen states. Although this point is beyond the present
article there are no conceptual difficulties to realize this step.

The weak points the advocated minimal model are more demanding.
The persisting challenges for theory are two-fold: (i) the high energy part of the spectrum at 
around $\approx$27\,meV is not reproduced and (ii) the down-bending behavior of the two 
lowest modes around $k\approx$0.75\,r.l.u. is not described properly. 
We presume that both discrepancies are due to the neglect of many-triplon
states in the present theory. 

Thus, an improved approach must be extended to states with more triplons.
A first step has been performed recently by Plumb {et al.} \cite{plumb15}
who applied diagrammatic perturbation theory to the hardcore triplons.
So far, none of the above stated challenges has been solved.
Thus the magnetic excitations in BCPO continue to be of great interest
because sizable anisotropic exchange couplings open fascinating 
routes to unconventional physics in quantum magnets \cite{romha15}.

\begin{acknowledgments}
We thank Young-June Kim, Kemp Plumb and Christian R\"uegg for providing data
and useful discussions. We are also grateful to
Michael Lang, Bruce Normand and Bernd Wolf for enriching comments.
We thank the Helmholtz Virtual Institute ``New states of matter and their excitations''
for partial financial support.
\end{acknowledgments}



\appendix

\section{Symmetry analysis of $\mathbf{D}_{1}$}
\label{chap_symmetry_D1}
The analysis of the vector $\mathbf{D}_{1}$ concerning the NN bonds, see 
Fig.\ \ref{pic_spinladder_dm} is demonstrated in detail to provide a complete presentation of the symmetry analysis. 

By applying the rotation RS$_{y}$ we map the bonds of the vectors $D_{1,LU}$and
 $D_{1,LO}$, respectively, to the bonds to which the vectors $D_{1,RO}$ and $D_{1,RU}$ belong.
 It is not necessary to rearrange the spin operators according to our notation
after the rotation because the  spin operators stay in the same order with regard of the $y$-coordinate. In this way, we obtain the following relations
\begin{subequations}
\begin{align}
\label{RSbD1LU}
\mathbf{D}_{1,RO}&=\text{RS}_{y}\left(\mathbf{D}_{1,LU}\right)\\
\mathbf{D}_{1,RU}&=\text{RS}_{y}\left(\mathbf{D}_{1,LO}\right)\\
\mathbf{D}_{1,LU}&=\text{RS}_{y}\left(\mathbf{D}_{1,RO}\right)\\
\mathbf{D}_{1,LO}&=\text{RS}_{y}\left(\mathbf{D}_{1,RU}\right).
\end{align}
\end{subequations}
Second, we consider the rotation R$_{x}$ and obtain 
\begin{subequations}
\begin{align}
\label{RaD1LU}
\mathbf{D}_{1,RO}&=-\text{R}_{x}\left(\mathbf{D}_{1,LU}\right)\\
\mathbf{D}_{1,RU}&=-\text{R}_{x}\left(\mathbf{D}_{1,LO}\right)\\
\mathbf{D}_{1,LU}&=-\text{R}_{x}\left(\mathbf{D}_{1,RO}\right)\\
\label{RaD1RU}
\mathbf{D}_{1,LO}&=-\text{R}_{x}\left(\mathbf{D}_{1,RU}\right). 
\end{align}
\end{subequations}
After the rotation R$_{x}$ the spin operators have to be swapped to comply
with our convention. Thus an additional minus sign appears in 
Eqs.\ \eqref{RaD1LU}-\eqref{RaD1RU}. 

Next, the reflection S$_{xy}$ is applied yielding
\begin{subequations}
\begin{align}
\label{SabD1LU}
\mathbf{D}_{1,RU}&=-\text{S}_{xy}\left(\mathbf{D}_{1,LU}\right)\\
\mathbf{D}_{1,RO}&=-\text{S}_{xy}\left(\mathbf{D}_{1,LO}\right)\\
\mathbf{D}_{1,LO}&=-\text{S}_{xy}\left(\mathbf{D}_{1,RO}\right)\\
\mathbf{D}_{1,LU}&=-\text{S}_{xy}\left(\mathbf{D}_{1,RU}\right).
\end{align}
\end{subequations}
The additional minus sign occurs due to the pseudovector properties of the spin operators. 

Now we derive the relations between the vectors 
$\mathbf{D}_{1}$ which arise from applying the reflection S$_{xz}$
\begin{subequations}
\begin{align}
\label{SacD1LU}
\mathbf{D}_{1,LO}&=\text{S}_{xz}\left(\mathbf{D}_{1,LU}\right)\\
\mathbf{D}_{1,LU}&=\text{S}_{xz}\left(\mathbf{D}_{1,LO}\right)\\
\mathbf{D}_{1,RU}&=\text{S}_{xz}\left(\mathbf{D}_{1,RO}\right)\\
\mathbf{D}_{1,RO}&=\text{S}_{xz}\left(\mathbf{D}_{1,RU}\right).
\end{align}
\end{subequations}
In this case the minus signs resulting from the pseudovector properties and the rearrangement of the spin operators compensate. 

Finally, we employ the reflection SS$_{yz}$ to receive the following relations
\begin{subequations}
\begin{align}
\label{SbcD1LU}
\mathbf{D}_{1,LO}&=-\text{SS}_{yz}\left(\mathbf{D}_{1,LU}\right)\\
\mathbf{D}_{1,LU}&=-\text{SS}_{yz}\left(\mathbf{D}_{1,LO}\right)\\
\mathbf{D}_{1,RU}&=-\text{SS}_{yz}\left(\mathbf{D}_{1,RO}\right)\\
\mathbf{D}_{1,RO}&=-\text{SS}_{yz}\left(\mathbf{D}_{1,RU}\right).  
\end{align}
\end{subequations}
Here the minus sign occurs because of the pseudovector properties of the spin operators.

With the above relations we are now able to derive the parity and the behavior of the sign along the legs of the ladder of the vector $\mathbf{D}_{1}$. 
We start from the ansatz 
\begin{equation}
 \mathbf{D}_{1,LU}=c_{x}\mathbf{e}_{x}+c_{y}\mathbf{e}_{y}+c_{z}\mathbf{e}_{z},
\end{equation}
which means that $\mathbf{D}_{1,LU}$ is an 
arbitrary combination of the unit vectors $\mathbf{e}_{x}$, $\mathbf{e}_{y}$ and 
$\mathbf{e}_{z}$ with constant real coefficients $c_{x}$, $c_{y}$ and $c_{z}$. 
Using this ansatz in \eqref{SacD1LU} we obtain
\begin{equation}
\label{D1LOwithSac}
\mathbf{D}_{1,LO}=c_{x}\mathbf{e}_{x}-c_{y}\mathbf{e}_{y}+c_{z}\mathbf{e}_{z}. 
\end{equation}
Additionally, we insert the ansatz in  \eqref{SbcD1LU} 
and obtain 
\begin{equation}
\label{D1LOwithSbc}
 \mathbf{D}_{1,LO}=c_{x}\mathbf{e}_{x}-c_{y}\mathbf{e}_{y}-c_{z}\mathbf{e}_{z}.
\end{equation}
To fulfill Eqs.\ \eqref{D1LOwithSac} and \eqref{D1LOwithSbc},
 the $z$-component has to vanish, $c_{z}=0$. 
Using Eqs.\ \eqref{SabD1LU} and \eqref{RSbD1LU}, respectively, we obtain 
\begin{subequations}
 \begin{align}
  \mathbf{D}_{1,RU}&=-c_{x}\mathbf{e}_{x}-c_{y}\mathbf{e}_{y}\\
  \mathbf{D}_{1,RO}&=-c_{x}\mathbf{e}_{x}+c_{y}\mathbf{e}_{y}.
 \end{align}
\end{subequations}

In conclusion, we see that the sign of the $x$-component does not change along the legs, i.e.,
 the signs of the $x$-component of the vectors 
$\mathbf{D}_{1,LO}$ and $\mathbf{D}_{1,LU}$ are the same as the signs of the 
$x$-component of the vectors $\mathbf{D}_{1,RO}$ and $\mathbf{D}_{1,RU}$.

In contrast, the $y$-component alternates along the legs, i.e., 
the signs of the $y$-component of the vectors $\mathbf{D}_{1,LO}$ and 
$\mathbf{D}_{1,LU}$ differ, so do the signs of the vectors $\mathbf{D}_{1,RO}$ and 
$\mathbf{D}_{1,RU}$. To determine the parity of 
$\mathbf{D}_{1}$ we  compare the sign of each components of $\mathbf{D}_{1,LO}$ with 
then one of $\mathbf{D}_{1,RO}$ and $\mathbf{D}_{1,LU}$ with $\mathbf{D}_{1,RU}$. 
As a result we find that the components on the left leg have a different sign 
than the components on the right leg. Hence, the parity of $\mathbf{D}_{1}$ is odd.

\section{Symmetry analysis of $\mathbf{D}_{0}$}
\label{chap_symmetry_D0}

To determine the direction of the vector $\mathbf{D}_{0}$
determining the DM-term on the rungs, see Fig.\ \ref{pic_spinladder_dm}, 
the third selection rule of Moriya \cite{moriy60b} is applied. This rule indicates that 
$\mathbf{D}_{0}$ has to point into the $y$-direction due to the existing 
symmetry S$_{xz}$. To analyze the behavior of the sign along the legs we 
use RS$_{y}$ or SS$_{yz}$ yielding
\begin{equation}
\mathbf{D}_{0,U}=-\mathbf{D}_{0,O},
\end{equation}
which means that the sign of $\mathbf{D}_{0}$ alternates along the legs.

\section{Symmetry analysis of $\mathbf{D}_{2}$}
\label{chap_symmetry_D2}

The analysis of the vector $\mathbf{D}_{2}$ concerning the NNN bonds, see Fig.\ 
\ref{pic_spinladder_dm}, is more complicated, but analogous to 
the symmetry analysis of $\mathbf{D}_{1}$ in App. \ref{chap_symmetry_D1}.
By applying the rotation RS$_{y}$ we obtain the following relations
\begin{subequations}
\begin{align}
\mathbf{D}_{2,LU}&=\text{RS}_{y}\left(\mathbf{D}_{2,RO}\right)\\
\mathbf{D}_{2,LO}&=\text{RS}_{y}\left(\mathbf{D}_{2,RU}\right)\\
\mathbf{D}_{2,RU}&=\text{RS}_{y}\left(\mathbf{D}_{2,LO}\right)\\
\mathbf{D}_{2,RO}&=\text{RS}_{y}\left(\mathbf{D}_{2,LU}\right).
\end{align}
\end{subequations}
Considering the rotation R$_{x}$ yields
\begin{subequations}
\begin{align}
\label{RxD2RU}
\mathbf{D}_{2,LU}&=-\text{R}_{x}\left(\mathbf{D}_{2,RU}\right)\\
\mathbf{D}_{2,LO}&=-\text{R}_{x}\left(\mathbf{D}_{2,RO}\right)\\
\mathbf{D}_{2,RU}&=-\text{R}_{x}\left(\mathbf{D}_{2,LU}\right)\\
\label{RxD2LO}
\mathbf{D}_{2,RO}&=-\text{R}_{x}\left(\mathbf{D}_{2,LO}\right).
\end{align}
\end{subequations}
After the rotation R$_{x}$ the spin operators have to be rearranged to 
conserve the convention regarding the sequence of $y$-coordinates. 
This is the reason for the minus signs in Eqs.\ (\ref{RxD2RU}-\ref{RxD2LO}). 

Next we apply the reflection S$_{xy}$ from where we find
\begin{subequations}
\begin{align}
\mathbf{D}_{2,LU}&=-\text{S}_{xy}\left(\mathbf{D}_{2,RU}\right)\\
\mathbf{D}_{2,LO}&=-\text{S}_{xy}\left(\mathbf{D}_{2,RO}\right)\\
\label{SxyD2LU}
\mathbf{D}_{2,RU}&=-\text{S}_{xy}\left(\mathbf{D}_{2,LU}\right)\\
\mathbf{D}_{2,RO}&=-\text{S}_{xy}\left(\mathbf{D}_{2,LO}\right).
\end{align}
\end{subequations}
The additional minus sign occurs because of the pseudovector properties of the spin operators. 
Then we derive the relations between the vectors 
$\mathbf{D}_{2}$  arising from applying the reflection S$_{xz}$
\begin{subequations}
\begin{align}
\label{SxzD2LU}
\mathbf{D}_{2,LU}&=\text{S}_{xz}\left(\mathbf{D}_{2,LU}\right)\\
\mathbf{D}_{2,LO}&=\text{S}_{xz}\left(\mathbf{D}_{2,LO}\right)\\
\mathbf{D}_{2,RU}&=\text{S}_{xz}\left(\mathbf{D}_{2,RU}\right)\\
\mathbf{D}_{2,RO}&=\text{S}_{xz}\left(\mathbf{D}_{2,RO}\right). 
\end{align}
\end{subequations}
In this case, the minus sign from the pseudovector properties and from
the rearrangement of the spin operators compensate. 

Finally, we use the reflection SS$_{yz}$ to derive the following relations
\begin{subequations}
\begin{align}
\mathbf{D}_{2,LU}&=-\text{SS}_{yz}\left(\mathbf{D}_{2,LO}\right)\\
\label{SSyzD2LU}
\mathbf{D}_{2,LO}&=-\text{SS}_{yz}\left(\mathbf{D}_{2,LU}\right)\\
\mathbf{D}_{2,RU}&=-\text{SS}_{yz}\left(\mathbf{D}_{2,RO}\right)\\
\label{SSyzD2RU}
\mathbf{D}_{2,RO}&=-\text{SS}_{yz}\left(\mathbf{D}_{2,RU}\right).
\end{align}
\end{subequations}
Here the minus sign appears due to the pseudovector properties of the spin operators.

As illustrated in Sect.\ \ref{symmetriesDcomponents} for $\mathbf{D}_{1}$
one can use the above relations to determine the behavior of the sign along the 
legs of the ladder and the parity of each $\mathbf{D}_{2}$-component. To this end, 
we make the ansatz 
\begin{equation}
\mathbf{D}_{2,LU}=d_{x}\mathbf{e}_{x}+d_{y}\mathbf{e}_{y}+d_{z}\mathbf{e}_{z},
\end{equation}
with real constant coefficients $d_{x}$, $d_{y}$ and $d_{z}$. Inserting this ansatz in
\eqref{SxzD2LU} we see that the $y$-component has to vanish.  This holds 
also for all other $\mathbf{D}_{2}$ vectors. From \eqref{SSyzD2LU} we obtain 
\begin{equation}
\mathbf{D}_{2,LO}=d_{x}\mathbf{e}_{x}-d_{z}\mathbf{e}_{z}.
\end{equation}
Using\eqref{SxyD2LU} yields
\begin{equation}
\mathbf{D}_{2,RU}=-d_{x}\mathbf{e}_{x}+d_{z}\mathbf{e}_{z}.
\end{equation}
Using this result and \eqref{SSyzD2RU} we obtain
\begin{equation}
 \mathbf{D}_{2,RO}=-d_{x}\mathbf{e}_{x}-d_{z}\mathbf{e}_{z}.
\end{equation}
As a conclusion, we find that the sign of the $x$-component does not change along the legs, 
i.e., the signs of the $x$-component of the vectors $\mathbf{D}_{2,LO}$ and $\mathbf{D}_{2,LU}$
are the same as the signs of the $x$-component of the vectors $\mathbf{D}_{2,RO}$ 
and $\mathbf{D}_{2,RU}$. In contrast, the $z$-component alternates along the legs, i.e.,
the sign of the $z$-component of the vectors $\mathbf{D}_{2,LO}$ and $\mathbf{D}_{2,LU}$ differ.
So do the signs of the vectors $\mathbf{D}_{2,RO}$ and $\mathbf{D}_{2,RU}$. 

Concerning the parity, we see that the parity of the $x$-component is odd, i.e., the signs 
of the $x$-components of $\mathbf{D}_{2,LU}$ and 
$\mathbf{D}_{2,RU}$ differ, so as the signs of the $x$-components of $\mathbf{D}_{2,LO}$ and 
$\mathbf{D}_{2,RO}$. Looking at the parity of the $z$-component we see that it is even, i.e.,
the sign of the $z$-components of $\mathbf{D}_{2,LU}$ and $\mathbf{D}_{2,RU}$ 
is the same, so as the corresponding signs in $\mathbf{D}_{2,LO}$ and $\mathbf{D}_{2,RO}$.

\section{Transformed anisotropic interaction terms}
\label{transformed_anisotropic_terms}

For completeness, we list all the transformed anisotropic interaction terms which do not vanish due to symmetry arguments and which are not listed in the main text
\begin{subequations}
\begin{align}
\nonumber
\mathcal{H}_{\mathrm{rung},\alpha\alpha}^{\Gamma,\mathrm{eff}}&=
-\Gamma_{0}^{\alpha\alpha}\sum_{k}a^2\left(k\right)
\\
&\left(t_{k}^{\alpha,\dagger}t_{-k}^{\alpha,\dagger}+
2t_{k}^{\alpha,\dagger}t_{k}^{\alpha}+t_{k}^{\alpha}t_{-k}^{\alpha}\right)
\\
\nonumber
\mathcal{H}_{\mathrm{NN},\alpha\alpha}^{\Gamma,\mathrm{eff}}&=
2\Gamma_{1}^{\alpha\alpha}\sum_{k}a^2\left(k\right)\cos\left(k\right)
\\
&\left(t_{k}^{\alpha,\dagger}t_{-k}^{\alpha,\dagger}+
2t_{k}^{\alpha,\dagger}t_{k}^{\alpha}+t_{k}^{\alpha}t_{-k}^{\alpha}\right)
\\
\nonumber
\mathcal{H}_{\mathrm{NN},xy}^{\Gamma,\mathrm{eff}}&=
-2\Gamma_{1}^{xy}\sum_{k}a\left(k\right)a\left(k+\pi\right)
\\
&\left(\mathrm{e}^{\mathrm{i}k}t_{k}^{x,\dagger}
\left(t_{-k-\pi}^{y,\dagger}+t_{k+\pi}^{y}\right)+\mathrm{h.c.}\right)
\\
\nonumber
\mathcal{H}_{\mathrm{NN},yx}^{\Gamma,\mathrm{eff}}&=
2\Gamma_{1}^{yx}\sum_{k}a\left(k\right)a\left(k+\pi\right)
\\
&\left(\mathrm{e}^{-\mathrm{i}k}t_{k}^{x,\dagger}
\left(t_{-k-\pi}^{y,\dagger}+t_{k+\pi}^{y}\right)+\mathrm{h.c.}\right)
\\
\nonumber
\mathcal{H}_{\mathrm{NNN},\alpha\alpha}^{\Gamma,\mathrm{eff}}&=
2\Gamma_{2}^{\alpha\alpha}\sum_{k}a^2\left(k\right)\cos\left(2k\right)
\\
&\left(t_{k}^{\alpha,\dagger}t_{-k}^{\alpha,\dagger}+
2t_{k}^{\alpha,\dagger}t_{k}^{\alpha}+t_{k}^{\alpha}t_{-k}^{\alpha}\right)
\end{align}
\end{subequations}
whereas $\alpha\in\{x,y,z\}$.

\section{Precise form of the 8$\times$8 matrix}
\label{concrete_8x8matrix}
The complete 8$\times$8 commutation matrix has the structure
\begin{equation}
\mathcal{M}_{\mathrm{all},xy,J_{2}}=\begin{pmatrix}
                                     \mathcal{M}_{xx} & \mathcal{M}_{xy}\\
                                     \mathcal{M}_{yx} & \mathcal{M}_{yy}
                                    \end{pmatrix},
\end{equation}
where each entry denotes a 4$\times$4 matrix. 
The matrix $\mathcal{M}_{xx}$ has the following form
\begin{equation}
\mathcal{M}_{xx}=\begin{pmatrix}
                  A_{\omega} & J_{2} & -A\left(k\right) & -J_{2}\\
		  J_{2} & A_{\omega2} & -J_{2} & -A\left(k+\pi\right)\\
		  A\left(k\right) & J_{2} & -A_{\omega} & - J_{2}\\
		  J_{2} & A\left(k+\pi\right) & -J_{2} & -A_{\omega2}
                 \end{pmatrix}.
\end{equation}
Here we used the abbreviation
\begin{equation}
A_{\omega2}\coloneqq \omega_{2}+A\left(k+\pi\right).
\end{equation}
The expressions for $A_{\omega}$, $J_{2}$, $\omega_{2}$ and $A\left(k\right)$ can be found 
in Eqs.\ \eqref{Aomega}, \eqref{J_2}, \eqref{omega2}, and 
\eqref{Ak}.

The matrix $\mathcal{M}_{yy}$ is given by
\begin{equation}
\mathcal{M}_{yy}=\begin{pmatrix}
                  E_{\omega} & J_{2} & -E\left(k\right) & -J_{2}\\
		  J_{2} & E_{\omega2} & -J_{2} & -E\left(k+\pi\right)\\
		  E\left(k\right) & J_{2} & -E_{\omega} & -J_{2}\\
		  J_{2} & E\left(k+\pi\right) & -J_{2} & -E_{\omega2}
                 \end{pmatrix}.
\end{equation}
The introduced coefficients are given by
\begin{subequations}
 \begin{align}
  E_{\omega}&\coloneqq\omega_{1}+E\left(k\right)\\
E_{\omega2}&\coloneqq\omega_{2}+E\left(k+\pi\right)\\
E\left(k+\pi\right)&\coloneqq d_{2}+\Gamma_{0}^{yy}\left(k\right)+
\Gamma_{1}^{yy}\left(k\right)+\Gamma_{2}^{yy}\left(k\right)\\
E\left(k\right)&\coloneqq d_{1}+\Gamma_{0}^{yy}\left(k+\pi\right)+
\Gamma_{1}^{yy}\left(k+\pi\right)+\Gamma_{2}^{yy}\left(k+\pi\right).
 \end{align}
\end{subequations}
The exact form of the abbreviations for $\omega_{1}$, $d_{2}$, $d_{1}$, 
$\Gamma_{0}^{yy}\left(k\right)$, $\Gamma_{1}^{yy}\left(k\right)$, 
and $\Gamma_{2}^{yy}\left(k\right)$ are shown in Eqs.\ \eqref{omega1}, \eqref{d2}, \eqref{d1},
 \eqref{Gamma0yyk}, \eqref{Gamma1yyk}, and \eqref{Gamma2yyk}.

The entries of the matrix $\mathcal{M}_{xy}$ read
\begin{equation}
\mathcal{M}_{xy}=\begin{pmatrix}
                  \mathrm{i}D_{1}^{z}\left(k\right) & F_{-,+} & -
									\mathrm{i}D_{1}^{z}\left(k\right) & F_{+,-}\\
		  F_{+,+} & \mathrm{i}D_{1}^{z}\left(k+\pi\right) & F_{-,-} & -
			\mathrm{i}D_{1}^{z}\left(k+\pi\right)\\
		  \mathrm{i}D_{1}^{z}\left(k\right) & F_{-,+} & 
			-\mathrm{i}D_{1}^{z}\left(k\right) & F_{+,+}
			\\
		  F_{+,+} & \mathrm{i}D_{1}^{z}\left(k+\pi\right) & F_{-,-} & 
			-\mathrm{i}D_{1}^{z}\left(k+\pi\right)
                 \end{pmatrix}.
\end{equation}
Here we used the abbreviations
\begin{subequations}
\begin{align}
D_{1}^{z}\left(k\right)&\coloneqq 4D_{1}^{z}a^2\left(k\right)\sin\left(k\right)
\\
F_{\pm,\pm}\left(k\right)&\coloneqq\pm\mathrm{i}\Gamma_{1}^{xy}\left(k\right)
\pm\mathrm{i}D_{2}^{z}\left(k\right).
\end{align}
\end{subequations}
The expressions for $\Gamma_{1}^{xy}\left(k\right)$ and $D_{2}^{z}\left(k\right)$ are given in
Eqs.\ \eqref{Gamma1xyk} and \eqref{D2zk}.

The last matrix $\mathcal{M}_{yx}$ has the form
\begin{equation}
 \mathcal{M}_{yx}=\begin{pmatrix}
                   -\mathrm{i}D_{1}^{z}\left(k\right) & F_{-,-} & 
									\mathrm{i}D_{1}^{z}\left(k\right) & F_{+,+}\\
		    F_{+,-} & -\mathrm{i}D_{1}^{z}\left(k+\pi\right) & F_{-,+} & 
				\mathrm{i}D_{1}^{z}\left(k+\pi\right)\\
		    -\mathrm{i}D_{1}^{z}\left(k\right) & F_{-,-} & 
				\mathrm{i}D_{1}^{z}\left(k\right) & F_{+,+}\\
		    F_{+,-} & -\mathrm{i}D_{1}^{z}\left(k+\pi\right) & F_{-,+} & 
				\mathrm{i}D_{1}^{z}\left(k+\pi\right)
                  \end{pmatrix}.
\end{equation}

\end{document}